\documentclass[12pt]{iopart}
\usepackage[utf8]{inputenc}
\usepackage{booktabs}
\usepackage{url}
\usepackage{cite}
\usepackage{graphicx}
\usepackage{placeins}
\usepackage[dvipsnames,table]{xcolor}
\definecolor{tabblue}{RGB}{31,119,180}
\definecolor{tabred}{RGB}{214,39,40}
\definecolor{tabpurple}{RGB}{148,103,189}
\definecolor{taborange}{RGB}{255,127,14}
\definecolor{tabpink}{RGB}{227,119,194}
\expandafter\let\csname equation*\endcsname\relax
\expandafter\let\csname endequation*\endcsname\relax
\usepackage{microtype}
\usepackage{amsmath}
\usepackage{amssymb}
\usepackage{bm}

\newcommand{\dbp}{\delta \! B_{\parallel}}
\let\Pi\varPi

\providecommand{\kperp}{k_\perp}

\makeatletter
\long\def\@makecaption#1#2{\vskip\abovecaptionskip
  \normalsize\rm
  \sbox\@tempboxa{{\bf #1.} #2}%
  \ifdim \wd\@tempboxa >\hsize
    {\bf #1.} #2\par
  \else
    \hb@xt@\hsize{\hfil\box\@tempboxa\hfil}%
  \fi
  \vskip\belowcaptionskip
  \def\@tempa{table}\ifx\@captype\@tempa \vskip\tablecaptionskip \fi}
\let\@portraitcaption=\@makecaption
\newlength\tablecaptionskip
\makeatother

\begin{document}

\title[Towards reactor-relevant electromagnetic turbulence in MAST-U]{Towards reactor-relevant electromagnetic turbulence in the MAST-U spherical tokamak}

\author{D. Kennedy$^{1}$, Y. Zhang$^{1}$, B.S. Patel$^{1}$\footnote{Present address: Gridfire Inc, 3422 Old Capitol Trail, Suite 700, New Castle, Wilmington 19808-6124 Delaware, USA}, S. Blackmore$^{1}$, C. Vincent$^{1}$, C.M. Roach$^{1}$}

\address{$^1$UKAEA (United Kingdom Atomic Energy Authority), Culham Campus, Abingdon,
Oxfordshire, OX14 3DB, UK}

\ead{daniel.kennedy@ukaea.uk}

\begin{abstract}
We investigate electromagnetic ion-scale turbulence in a high-performing H-mode plasma of the Mega Ampere Spherical Tokamak Upgrade (MAST-U) with local gyrokinetic simulations, asking whether it approaches the non-zonal transition: the nonlinear transition to a high-flux state above a critical value of $q^{2}\beta_{e}$ found in simulations of STEP~\cite{Zhang2025b, kennedy2026}, where $q$ is the safety factor and $\beta_{e}$ the ratio of electron thermal pressure to magnetic pressure. We examine two flux surfaces: one at $t = 531\,\mathrm{ms}$, close to peak normalised pressure $\beta_N$, at normalised poloidal flux $\psi_N = 0.50$; and one at $t = 591\,\mathrm{ms}$ and $\psi_N = 0.60$, the ideal-ballooning-stable surface of largest $q^{2}\beta_e$ inside $\psi_N = 0.8$ at that time. On both surfaces the dominant ion-scale instabilities are microtearing and electromagnetic ballooning-like modes, as in STEP. Without equilibrium flow shear, at least one of the surfaces lies above the large flux threshold with every set of kinetic species we simulate, with ion-scale heat fluxes far exceeding power balance, in turbulence characterised by a highly stochastic magnetic field. Strong flow shear driven by neutral-beam injection in the real experiment suppresses this turbulence and holds the plasma below the transition. When the carbon impurity is included in the simulation, the ion-scale flux is then close to power balance on both surfaces. Without rotation, gyrokinetics suggests the transition would limit the accessible profiles.  This could degrade MAST-U plasma performance using heating schemes that lack a momentum source, and this will be tested in forthcoming electron Bernstein wave heating experiments. In the MAST-U experiment, large-scale MHD, the loss of rotation and the loss of performance coincide; a local model cannot resolve the role of the coherent low-$n$ activity, which motivates the need for future work using hybrid magnetohydrodynamic--gyrokinetic or global gyrokinetic models.
\end{abstract}

\vspace{2pc}

\section{Introduction}
\label{sec:intro}

The Mega Ampere Spherical Tokamak Upgrade (MAST-U) is a spherical tokamak (ST) at the Culham Campus of the UK Atomic Energy Authority~\cite{Harrison2019, harrison2026}. Compared with the Mega Ampere Spherical Tokamak (MAST)~\cite{Chapman2015}, it has a higher toroidal magnetic field, plasma current, and pulse length, and an additional off-axis neutral-beam injector (NBI)~\cite{Morris2018, Harrison2019}. Its physics results are important to the design of future STs~\cite{harrison2026}.

The ST concept~\cite{Peng1986} is a promising candidate for a compact fusion power plant~\cite{Wilson2004, Menard2016}. Its low aspect ratio allows a high plasma $\beta$, the ratio of plasma to magnetic pressure, at modest toroidal field. The UK STEP (Spherical Tokamak for Energy Production) Fusion programme~\cite{meyer2024, STEP, meyer2022, Chapman2024} aims to develop a compact power plant generating over $100\,\mathrm{MW}$ of net electric power, based on the ST concept~\cite{Waldon2024}. In its first phase, the programme has focused on designing reference plasma equilibria using simplified core-transport models, with the anomalous (turbulent) transport prescribed by a rescaled Bohm/gyro-Bohm model tuned on results from MAST~\cite{Erba1997, tholerus2024}. 

Future ST power plants such as STEP are expected to operate at large values of $\beta_e$, where $\beta_e = 2\mu_0 n_e T_e/B_0^2$ is the ratio of electron pressure to magnetic pressure, $n_e$ and $T_e$ are the electron density and temperature, and $B_0$ is a reference magnetic field. In such plasmas, electromagnetic (EM) effects play a leading role, and first-principles, higher-fidelity models such as gyrokinetics (GK) are required to provide more accurate predictions of the transport. Of course, to build confidence in the predictive capability of such models they must be rigorously and extensively compared to experiments. Transport and confinement in spherical tokamak experiments is extensively reviewed in \cite{kaye2021}, and it has been demonstrated that electromagnetic turbulence plays a significant role in today's machines. The value of $\beta_e$ required in the core of STEP is somewhat larger than is typically observed on present-day machines such as MAST-U~\cite{tholerus2024, kennedy2023a}, but other potentially more significant differences expected in STEP include: much lower collisionality; elevated $q$; higher bootstrap fraction; and lower rotation. Recent work~\cite{kennedy2026} identifies a critical value of $\beta_e$ for the onset of strong electromagnetic turbulence in STEP, which is comparable to the values attained in present-day machines (while reference STEP equilibria are far in excess of this limit). In this work, we examine whether MAST-U is able to access turbulent regimes similar to future machines like STEP, and whether MAST-U can probe electromagnetic turbulence in high-performance plasmas.

\subsection{Electromagnetic instabilities and turbulence in STEP}
\label{sec:intro-step}

First GK analyses of STEP-like reference flat-top operating points~\cite{Patel2022,kennedy2023a} identified unstable \textit{hybrid} kinetic ballooning modes (hKBMs)---which combine features of kinetic ballooning modes (KBMs), ion-temperature-gradient-driven modes (ITGs), and trapped-electron modes (TEMs)---as the dominant instabilities, with subdominant microtearing modes (MTMs) also present~\cite{Patel2022, kennedy2023a, Giacomin2024, Giacomin_Dickinson_Dorland_Mandell_Bokshi_Casson_Dudding_Kennedy_Patel_Roach_2025}. The hKBMs are very sensitive to flow shear and are unstable only when the parallel magnetic-field fluctuation $\dbp$ is retained~\cite{kennedy2024}. Subsequent nonlinear simulations~\cite{Giacomin2024, Giacomin_Dickinson_Dorland_Mandell_Bokshi_Casson_Dudding_Kennedy_Patel_Roach_2025, kennedy2026} revealed that, in the absence of equilibrium flow shear, hKBM-driven turbulence can generate heat fluxes far exceeding the available heating. At a critical value of $\beta_e$ the heat flux increases abruptly by several orders of magnitude, whilst the plasma remains stable to the ideal ballooning mode (IBM). This is a nonlinear phenomenon, meaning that this transition is decoupled from the linear threshold of any instability~\cite{Pueschel2013, Rath2022}. Thus, following~\cite{Pueschel2013}, we refer to this transition as the \emph{non-zonal transition} (NZT). Global EM simulations including $\dbp$ also reveal clear evidence of a transition to extremely large heat fluxes~\cite{kennedy2026global}.

\subsection{Stress balance, the $q^{2}\beta_{e}$ threshold and the reverse transition}
\label{sec:intro-stress-balance}

Recent work~\cite{Zhang2025a, Zhang2025b,kennedy2026} finds that the non-zonal transition occurs when the magnetic-flutter stress extracts energy from zonal flows, key regulators of ITG turbulence~\cite{Dimits2000, Ivanov2020, Ivanov2022}, at a rate that exceeds the injection of energy from the $E\times B$ stress~\cite{Zhang2025b}, resulting in a regime with weak zonal flows and elevated heat fluxes. When the Alfv\'en frequency for the lowest parallel wavenumber mode becomes comparable to the magnetohydrodynamic (MHD) interchange growth rate, the net turbulent torque opposes the zonal flow~\cite{Zhang2025b, kennedy2026}. Equating these two timescales yields a critical beta,
\begin{equation}
  \beta_{e,\mathrm{crit}} \sim \frac{L_T}{q^2 R},
  \label{eq:beta_crit}
\end{equation}
where $q$ is the safety factor, $L_T$ is the temperature-gradient scale length and $R$ is the major radius. It was shown that in Cyclone Base Case~\cite{Dimits2000}, ST40~\cite{McNamara2024}, and STEP geometries that the non-zonal transition indeed occurs when $q^2\beta_e$ exceeds a critical threshold, $C_{\mathrm{nl}}$, below the threshold for ideal and kinetic ballooning mode instability. The value of $C_{\mathrm{nl}}$ is not universal, but depends on the device and equilibrium shaping. At larger values of the pressure gradient, the heat flux returns to comparatively low levels, through $\beta'$ stabilisation~\cite{bourdelle2003, bourdelle2005} and access to a second-stable regime beyond the IBM threshold~\cite{kennedy2026}. Between this regime and the zonally regulated state at low $q^2\beta_e$ lies a ``no-go zone'' of large EM fluxes. In an experiment with realistic sources, no plasma could sustain the fluxes of the no-go zone: its profiles would instead relax until $q^2\beta_e$ fell back to $C_{\mathrm{nl}}$, so that the non-zonal transition could act as a very stiff limit, pinning the profiles at a threshold that lies below the ideal $\beta$ limit. The IBM threshold in the plane of magnetic shear $\hat{s}$ and normalised pressure gradient $\alpha_{\mathrm{MHD}}$ of~(\ref{eq:shat_alpha_def}) is straightforward to calculate~\cite{connor1979, dewar1983, davies2022, Gaur_Buller_Ruth_Landreman_Abel_Dorland_2023}, and serves as a useful proxy for both edges of this zone~\cite{kennedy2026}. 

The non-zonal transition in this sense is an ion-scale phenomenon concerning the ability of ion-scale turbulence to build and sustain large-scale zonal flows.

\subsection{Turbulence in high-performance spherical tokamaks}
\label{sec:intro-st-turbulence}

In beam-heated high-confinement-mode (H-mode) plasmas in MAST and the National Spherical Torus Experiment (NSTX)~\cite{Ono2000}, the ion heat transport is often close to neoclassical, whereas the electron heat transport is anomalous~\cite{kaye2021}. The comparatively weak ion transport has been attributed in part to the strong flow shear driven by NBI~\cite{Roach2009, kaye2021, singh2026}. The streamer-like structures responsible for EM high-transport states above the NZT can be sheared by zonal flows or by radial gradients in the equilibrium rotation profile~\cite{kennedy2026}. Local GK studies of MAST equilibria have identified ion-scale MTMs and electron-scale electron-temperature-gradient (ETG) modes as possible sources of the electron heat transport~\cite{Applegate2004, Roach2005, Joiner2006, Roach2009, Colyer2017}. It has also recently been shown that microtearing turbulence in MAST can form stochastic layers~\cite{Giacomin2023microtearing}, and is less sensitive to equilibrium $E\times B$ shear than in NSTX~\cite{patel2025exb}, which might explain the observed anomalous electron heat transport in the presence of rotational shear.  Semi-empirical Rechester-Rosenbluth models can describe electron heat transport from stochastic fields, and transport calculations with such models have been shown to be in reasonable agreement with measurements from some MAST plasmas \cite{Palermo2025}.

A companion study by Patel~\textit{et al.}~\cite{patel2026etg} finds an example of one H-mode discharge (\#48657, see Section~\ref{sec:intro-48657}) where the electron-scale (ETG) transport is absent in the high performance phase, and only appears to seemingly approach the experimental transport level in the low-performance phase after the rotation collapse. It was suggested in~\cite{patel2026etg} that ion-scale simulations are required wherever ETG modes are stable. In the high-performance state, this MAST-U plasma has large $\beta_e,$ and at the surfaces studied in~\cite{patel2026etg} and here, the instabilities are predominantly ion-scale, with the electron scale stable or carrying a negligible flux (Sections~\ref{sec:linear-spectra} and~\ref{sec:conclusions-shortfall}).

\subsection{MAST-U discharge \#48657}
\label{sec:intro-48657}

In this paper, we present local electromagnetic ion-scale gyrokinetic simulations of the high-performance MAST-U discharge \#48657. Discharge \#48657 is an H-mode plasma reaching high normalised beta, $\beta_N = 3.9$ at $t = 531\,\mathrm{ms}$, close to peak-performance in MAST-U, with $3.4\,\mathrm{MW}$ of NBI~\cite{harrison2026, patel2026etg}. Here, $\beta_N \equiv \beta_t a B_T/I_p$, where $\beta_t$ is the toroidal beta, $a$ is the minor radius in metres, $B_T$ is the toroidal field in tesla and $I_p$ is the plasma current in MA, it measures the plasma pressure against the ideal MHD beta limit~\cite{Troyon1984}. 

\begin{figure}[tbp]
  \centering
  \begin{minipage}[t]{0.49\textwidth}
    \centering
    \includegraphics[width=\textwidth]{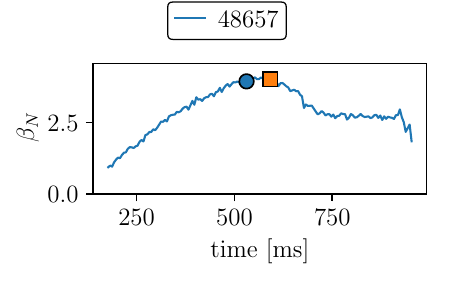}
  \end{minipage}\hfill
  \begin{minipage}[t]{0.49\textwidth}
    \centering
    \includegraphics[width=\textwidth]{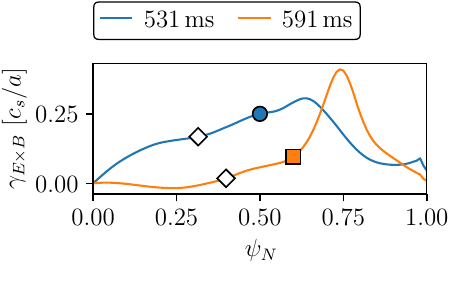}
  \end{minipage}
  \caption{MAST-U \#48657. $\beta_N$ against time (left-hand side); circle, S50 ($t = 531\,\mathrm{ms}$); square, S60 ($t = 591\,\mathrm{ms}$). $E\times B$ shearing rate against $\psi_N$ at those times (right-hand side); \textcolor{tabblue}{blue}, $531\,\mathrm{ms}$; \textcolor{taborange}{orange}, $591\,\mathrm{ms}$; open diamonds, the $q = 3/2$ surface at each time, inboard of the $q = 2$ surface of the $2/1$ neoclassical tearing mode (Section~\ref{sec:candidate-selection}); circle and square, the analysis surfaces S50 ($\psi_N = 0.50$) and S60 ($\psi_N = 0.60$) used in this work.}
  \label{fig:mast-profiles}
\end{figure}

As in most beam-heated MAST-U H-modes, confinement is limited by an $m/n = 2/1$ neoclassical tearing mode~\cite{harrison2026}, which reduces the core rotation at $t \approx 600\,\mathrm{ms}$ (Figure~\ref{fig:mast-profiles}). Here, $m$ and $n$ are the poloidal and toroidal mode numbers. Interestingly, the loss of core performance and the loss of shear rotation are very closely linked, suggesting that this discharge may be an excellent place to search for strong EM GK turbulence in present-day experiments, and try to understand its interaction with shear rotation. Which of the tearing mode, the loss of rotation and the loss of performance is cause and which is effect is a question to which we return in Section~\ref{sec:conclusions}. In interpretive modelling of \#48657 with the transport analysis code \texttt{TRANSP}~\cite{Pankin2025}, the anomalous electron heat diffusivity is an order of magnitude above the deuterium one during the high-$\beta_N$ phase, where ETG transport would be one candidate explanation~\cite{patel2026etg}. However, as previously remarked it was found that electron-scale transport under-predicts the anomalous transport seen by \texttt{TRANSP}. Furthermore, part of the $\beta_e$ profile at $t = 600\,\mathrm{ms}$ lies on the IBM boundary, where STEP-like conditions would be expected to give large fluxes due to EM ion-scale turbulence~\cite{kennedy2026}, and Harrison~\textit{et al.}~\cite{harrison2026} attribute this access to the flow shear driven by NBI. It is therefore interesting to explore whether such a plasma exists below the non-zonal transition, inside the no-go zone with its transport limited by flow shear, or in the second-stable regime, and whether ion-scale turbulence can account for the anomalous transport where the electron scale falls short.

In this work we examine two flux surfaces from separate times during the discharge: one close to peak $\beta_N$ (S50; $t = 531\,\mathrm{ms}$, $\psi_N = 0.50$) and one at the maximum IBM-stable $q^2\beta_e$ (S60; $t = 591\,\mathrm{ms}$, $\psi_N = 0.60$). These local equilibria (each a single flux surface at a single time) are chosen as detailed in Section~\ref{sec:candidate-selection}. The labels name the surfaces by $\psi_N$. At the time of peak $\beta_N$ we take the radius studied in~\cite{patel2026etg}, which lies on the stable inner edge of an ideal-ballooning-unstable pocket (Section~\ref{sec:candidate-surfaces}); the second surface is the ideal-ballooning-stable surface of largest $q^2\beta_e$ inside $\psi_N = 0.8$ at its time, in the band where the figure of merit of Section~\ref{sec:fom} is concentrated. Figure~\ref{fig:mast-profiles} shows $\beta_N$ against time and the $E\times B$ shearing rate (Section~\ref{sec:candidate-selection}) at the two times selected. Using local GK analysis we find that linear, electromagnetic, ion-scale instabilities (closely related to those anticipated in STEP) would be highly unstable in MAST-U if flow shear were weak or absent. Nonlinear simulations of these ion-scale instabilities find that, in the absence of equilibrium flow shear, both S50 and S60 sit inside the no-go zone (for $\gamma_E=0$) suggesting that they must be held there by the strong flow shear. When rotational effects at the experimental level are included in the simulations; both equilibria sit close to the power balance predicted by \texttt{TRANSP} modelling. Impurities and fast ions both play an important role in the nonlinear transport. 

The layout of this paper is as follows. Section~\ref{sec:candidate-selection} describes how the analysis times and flux surfaces are chosen. Section~\ref{sec:equilibrium-linear} discusses the local equilibrium, the ideal-ballooning stability and the linear stability of the two surfaces. Section~\ref{sec:nonlinear} presents the nonlinear simulations, and Section~\ref{sec:species} examines the role of impurities and fast ions, with two, three, and four kinetic species. Finally, our conclusions, together with predictions for the EBW-heated plasmas of the next MAST-U campaign, are presented in Section~\ref{sec:conclusions}.

\section{Selection of candidate times and flux surfaces}
\label{sec:candidate-selection}

In this section, we briefly discuss the selection criteria for the surfaces selected for detailed GK analysis. This is a difficult problem since we are searching for local equilibria that are as close as possible to peak-performance, where the turbulence is anticipated to be the most electromagnetic, whilst also trying to avoid activity from large-scale MHD modes that would invalidate the local GK ordering. It is important to emphasise that large-scale MHD cannot be avoided entirely in this discharge, and the selection below is a pragmatic best effort rather than a guarantee that the local model holds. We return to this issue in Section~\ref{sec:conclusions}.

\subsection{A figure of merit for ion-scale EM GK turbulence}
\label{sec:fom}

We aim to devise a figure of merit that can be used to identify surfaces and times within the discharge that are promising for GK analysis. The figure of merit is the product of four factors evaluated over $(\psi_N,\,t)$. 
The four factors combine to give a score
\begin{equation}
    S_{\mathrm{GK}}(\psi_N,t) =
        s_{\beta_N}\;  s_{\mathrm{drive}}\; G_{\mathrm{IBM}}\; s_{\mathrm{MHD}}^{1/2}.
    \label{eq:gk-score}
\end{equation}
with the factors themselves consisting of:

\begin{itemize}
    \item The pressure factor, $s_{\beta_N} \equiv \beta_N/\beta_N^{\max}$, with $\beta_N^{\max}$ the largest $\beta_N$ over the time window of the maps. This measures proximity to peak-performance.
    \item The drive factor, $s_{\mathrm{drive}} \equiv q^2\beta_e$, the control parameter of the stress-balance framework~\cite{Zhang2025b}, normalised by its largest IBM-stable value.
    \item A validity gate, $G_{\mathrm{IBM}}$, equal to one if the surface being scored is locally stable to the $n = \infty$ ideal ballooning mode and zero otherwise (Section~\ref{sec:fom-ibm}). 
    \item A magnetic activity factor $s_{\mathrm{MHD}}^{1/2}$ that penalises proximity to coherent large-scale MHD activity (Section~\ref{sec:fom-mhd}).
\end{itemize}

\begin{figure}[tbp]
  \centering
  \includegraphics[width=\textwidth]{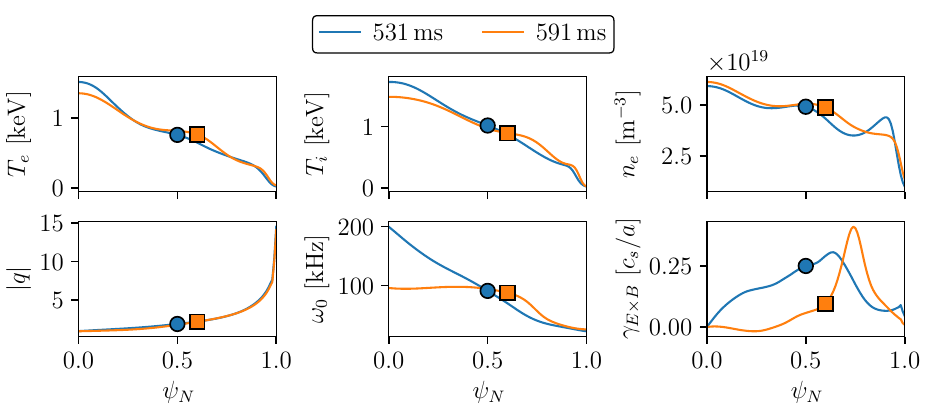}
  \caption{Profiles of \#48657 at $531\,\mathrm{ms}$ (\textcolor{tabblue}{blue}) and $591\,\mathrm{ms}$ (\textcolor{taborange}{orange}): electron temperature $T_e$, ion temperature $T_i$ and electron density $n_e$ (top row); $|q|$, rotation frequency $\omega_0$ and $E\times B$ shearing rate in $c_s/a$ (bottom row). Circle, S50 ($531\,\mathrm{ms}$, $\psi_N = 0.50$); square, S60 ($591\,\mathrm{ms}$, $\psi_N = 0.60$).}
  \label{fig:profile-snapshots}
\end{figure}

Figure~\ref{fig:profile-snapshots} shows the profiles at the two times ultimately selected by the analysis in this section. Here, $\gamma_{E\times B} = (r/q)\,\mathrm{d}\omega_0/\mathrm{d}r$ is the $E\times B$ shearing rate from the toroidal rotation alone, where $r$ is the local minor radius and $\omega_0$ is the toroidal angular rotation frequency. Rates are normalised to $c_s/a$, where $c_s = \sqrt{T_e/m_i}$ is the sound speed and $m_i$ is the deuterium mass. After peak $\beta_N$ the rotation flattens across the inner core, confining $\gamma_{E\times B}$ to the outer core, so the shearing rate at S50 is two to three times that at S60 (Figure~\ref{fig:mast-profiles}).

\subsubsection{Locating regions of IBM stability in time and radius}
\label{sec:fom-ibm}

The third factor~(\ref{eq:gk-score}) gates the score based on IBM stability. Local gyrokinetics requires stability to the $n=\infty$ IBM, the ideal MHD counterpart of the KBM, and so every analysis surface must be IBM-stable (Section~\ref{sec:equilibrium-linear}). This is ensured by the IBM validity gate, $G_{\mathrm{IBM}}$, which is applied locally, surface by surface; what a local IBM-unstable region implies for the global equilibrium is raised in Section~\ref{sec:candidate-surfaces}. The validity gate uses the eigenvalue $\gamma_{\mathrm{IBM}}$ of the ideal-ballooning equation, solved with \texttt{pyrokinetics}~\cite{pyrokinetics2024} on a Miller extended harmonic (MXH) fit to the local \texttt{TRANSP} equilibrium,
\begin{equation}
    G_{\mathrm{IBM}}(\psi_N,t) = \begin{cases} 1, & \gamma_{\mathrm{IBM}} \le 0,\\ 0, & \gamma_{\mathrm{IBM}} > 0,\end{cases}
    \label{eq:gibm-gate}
\end{equation}
where $\gamma_{\mathrm{IBM}}$ is the eigenvalue of the ideal ballooning equation, negative on the stable side, where ideal MHD has no growing mode, and positive on the unstable side. Pairings of radii and times that are IBM unstable are thus excluded.  

\subsubsection{Locating the coherent magnetic activity in radius}
\label{sec:fom-mhd}

The fourth factor appearing in~(\ref{eq:gk-score}), $s_{\mathrm{MHD}} \in (0,1]$, penalises coherent low-$n$ magnetic activity, since a growing mode perturbs the equilibrium that the local model holds fixed. Typically, experiments are never completely quiescent, and MAST-U hosts a wide range of MHD activity. As previously remarked, this is further complicated by this study intentionally seeking local equilibria likely to exhibit strong EM GK turbulence. We thus impose a soft gate on MHD activity: we score equilibria highly at times where there is no evidence of MHD activity at any radii, or where there is evidence of MHD activity but only at radii sufficiently far from the surface being scored, as set out in the fourth step below.

\begin{figure}[tbp]
  \centering
  \includegraphics{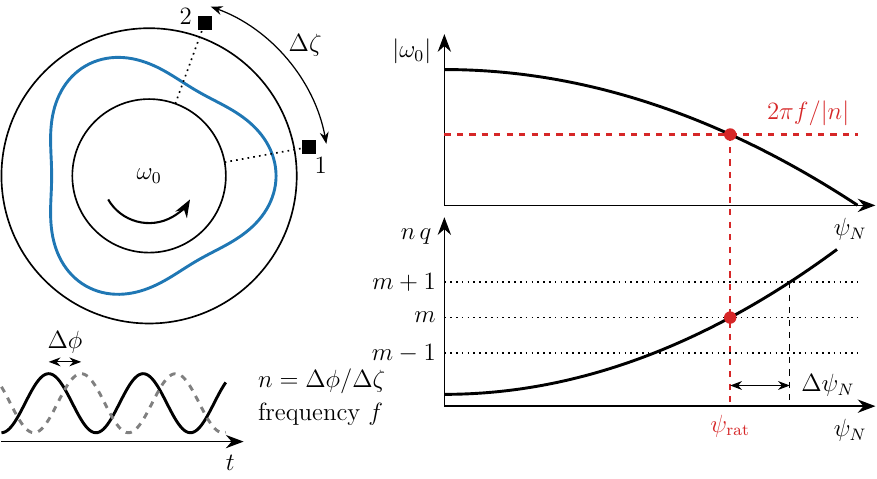}
  \caption{Schematic of the location of a coherent mode. Left-hand side, the torus from above, with a rotating $n = 3$ perturbation (\textcolor{tabblue}{blue}), the toroidal rotation $\omega_0$ and two outboard coils, 1 and 2 (squares), a toroidal angle $\Delta\zeta$ apart; the cross-phase $\Delta\phi$ between the coils gives $n$. Bottom left, the signals of coils 1 (solid) and 2 (dashed) against time $t$, offset by $\Delta\phi$. Right-hand side, $|\omega_0|$ (top) and $n\,q$ (bottom) against $\psi_N$. \textcolor{tabred}{Red}, the level $2\pi f/|n|$ of~(\ref{eq:mirnov-rational}) and the surface $\psi_{\mathrm{rat}}$ where $|\omega_0|$ reaches it; dotted, integer values of $n\,q$; $\Delta\psi_N$, the distance to the next resonance of the same $n$.}
  \label{fig:mirnov-cartoon}
\end{figure}

Mirnov coils detect magnetic fluctuations in the plasma through the voltage induced by the time-varying magnetic flux threading each coil~\cite{Hutchinson2002}. A coherent magnetic perturbation appears in these signals as a band of well-defined frequency $f$. Coils outside the plasma detect only coherent magnetic perturbations of low toroidal mode number, here $|n| = 1$, $2$ and $3$. The ion-scale turbulence in our simulations is broadband and peaks near $n = 20$, and its magnetic perturbation decays rapidly with distance from the surfaces on which it is driven, so it does not appear in these signals. We combine these measurements with the plasma rotation and safety-factor profiles to locate such structures in radius in three steps, and turn them into $s_{\mathrm{MHD}}$ in a fourth. Figure~\ref{fig:mirnov-cartoon} illustrates the procedure.

\begin{figure}[tbp]
  \centering
  \includegraphics[width=\textwidth]{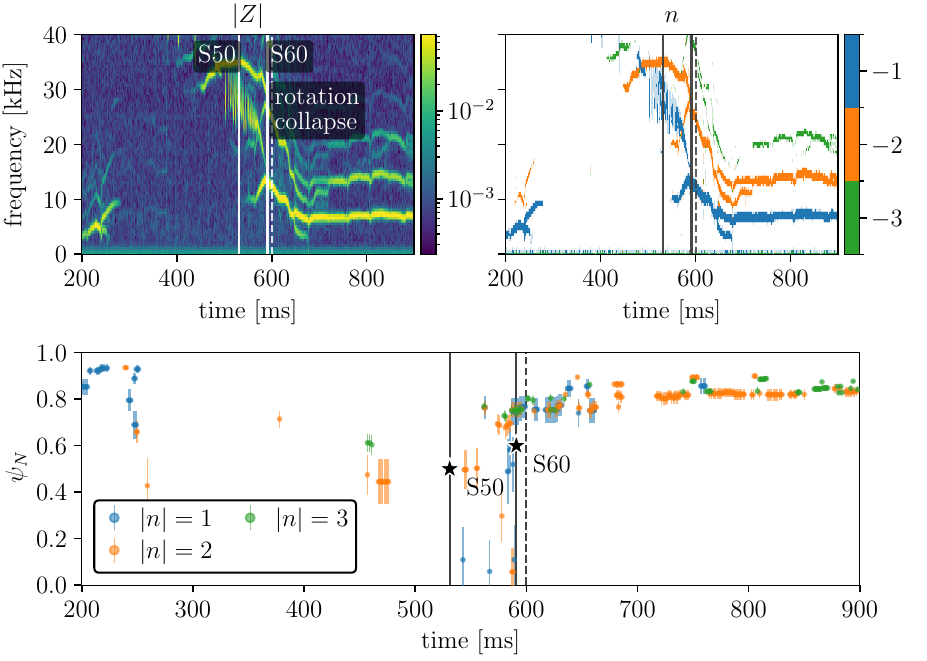}
  \caption{Mirnov cross-phase spectrograms of \#48657 from two toroidally separated outboard coils, $\Delta\zeta = 59.78^\circ$ apart. Short-time Fourier magnitude $|Z|$ (top left-hand side, note logarithmic scale). Toroidal mode number $n = \Delta\phi/\Delta\zeta$ of each time--frequency bin from the cross-phase $\Delta\phi$ between the two coils, searched over $n \in \{-3,\,-2,\,-1\}$, its sign giving the direction of propagation (top right-hand side); bins below the 95th percentile of $|Z|$ are masked (white). $\psi_{\mathrm{rat}}$ of~(\ref{eq:mirnov-rational}) against time for every band passing the frozen-in test $m = n\,q(\psi_{\mathrm{rat}})$, coloured by $|n|$, with bars spanning the distance to the next resonance of the same $n$ (bottom). Stars mark S50 and S60; solid vertical lines, the S50 and S60 analysis times ($531$ and $591\,\mathrm{ms}$); dashed vertical line, the rotation collapse ($600\,\mathrm{ms}$).}
  \label{fig:spectrogram-pair}
\end{figure}

Firstly, we find the toroidal mode numbers of coherent MHD modes. The toroidal mode number $n$ of a band follows from the cross-phase $\Delta\phi$ between two coils separated toroidally by an angle $\Delta\zeta$, since $\Delta\phi = n\,\Delta\zeta$. Our two coils are $\Delta\zeta = 59.78^\circ$ apart, which fixes $n$ only modulo six. Higher aliases would put the modes where the plasma rotates more slowly than anywhere inside $\psi_N = 0.995$, and so $|n| \in \{1,\,2,\,3\}$. Figure~\ref{fig:spectrogram-pair} shows a few such low-$|n|$ bands, whose frequency falls as the plasma slows towards peak performance and the rotation profile collapses (Figure~\ref{fig:profile-snapshots}, bottom row, middle column). Without an absolute calibration the coils give no amplitude at the mode, and so we do not calculate the island width.

Secondly, we find the location of surfaces in the plasma where the mode is active by matching the rotation of the mode to the rotation of the surfaces. A mode frozen into the plasma is carried past the coils by the local rotation, so that its frequency is the Doppler shift $2\pi f = |n|\,|\omega_0|$ at the surface on which it sits. We therefore locate the mode in radius, and penalise proximity to it, about the surface $\psi_{\mathrm{rat}}$ where
\begin{equation}
    |\omega_0(\psi_{\mathrm{rat}})| = \frac{2\pi f}{|n|}.
    \label{eq:mirnov-rational}
\end{equation}

Thirdly, we check that this is consistent with a rational safety factor. Since~(\ref{eq:mirnov-rational}) does not use $q$, we check each inferred location against the safety-factor profile. A resonant mode of poloidal mode number $m$ sits where $m = n\,q$, and so we require $n\,q(\psi_{\mathrm{rat}})$ to lie within $0.15$ of an integer $m$.

Fourthly, every frequency band placed in radius by the second step adds a penalty centred on its surface, and $s_{\mathrm{MHD}}$ is built from their sum,
\begin{equation}
  s_{\mathrm{MHD}}(\psi_N,t) = \exp\left[-3\sum_{\mathrm{bands}} \exp\left(-\frac{(\psi_N - \psi_{\mathrm{rat}})^2}{2\sigma^2}\right)\right],
  \qquad \sigma = 0.07,
  \label{eq:smhd}
\end{equation}
so that $s_{\mathrm{MHD}} = 1$ far from every band and about $0.05$ on one. A band counts whether or not it passes the check on $n\,q$, since an error in $q$ or $\omega_0$ can move a real mode off its rational surface; the check serves only to identify the resonant structures below. A band that cannot be placed in radius lowers $s_{\mathrm{MHD}}$ everywhere by a factor of $0.6$. Through the factor $s_{\mathrm{MHD}}^{1/2}$ in~(\ref{eq:gk-score}), a single mode halves the score of a surface within $\Delta\psi_N \approx 0.09$ of it, and reduces it to about a fifth on it. The width $\sigma$ stands in for the radial extent of the mode, which, as noted above, the coils cannot give. At the outboard midplane $\sigma$ is about $2\,\mathrm{cm}$, or $0.04\,a$, so that one mode affects a band about $5\,\mathrm{cm}$ across, comparable to the radial extents estimated below from the distance between resonances of the same $n$. Every band is given the same width and weight, whatever the size of its mode, which over-penalises a small mode and under-penalises a large one. This factor is therefore the most questionable part of the score: we justify it on grounds of pragmatism, and do not claim that it guarantees the absence of MHD effects at the selected surfaces.

At $t = 531\,\mathrm{ms}$, one of the times analysed in the companion study~\cite{patel2026etg}, no band passes this check. At $t = 591\,\mathrm{ms}$, three bands pass, corresponding to $m = 3$, $6$ and $9$, all near $\psi_N = 0.75$. These bands are consistent with a single structure on the $q = 3$ surface, seen at $n = 1$, $2$ and $3$. In the raw Thomson profiles around $591\,\mathrm{ms}$, $T_e$ at the $q = 3$ surface is flat in one of four frames and varies from frame to frame by up to four times the measurement error just inside it, near $\psi_N = 0.71$. At the radius of S60 it is steady, to within twice the error, so this variation lies outboard of the surface we analyse. The Thomson frames, $4$ to $6\,\mathrm{ms}$ apart, cannot resolve modes rotating at several kHz, so the profiles of Figure~\ref{fig:profile-snapshots} are consistent with, but do not demonstrate, a rotating $q = 3$ island. Over the discharge, about a quarter of the band--time samples pass the check (Figure~\ref{fig:spectrogram-pair}, bottom). After the rotation collapse, every mode we locate lies between $\psi_N = 0.75$ and $0.9$, outboard of both analysis surfaces. Before it, we also find $n = 2$ activity further in: near $\psi_N = 0.5$, close to S50, from $544$ to $556\,\mathrm{ms}$, and near $\psi_N = 0.7$ from $574$ to $586\,\mathrm{ms}$. We take the distance to the next resonance with the same $n$ as an upper limit on the radial extent of a structure. This gives about $\Delta\psi_N = 0.1$ for the $q = 3$ structure, with typical estimates about half this size; the bound is loose, and for $n = 1$ alone it would span much of the profile, but the structures located here are seen at $|n| = 2$ or $3$, which give the tighter estimates, so that the structure is several estimated widths away from the local calculation regions. At $531\,\mathrm{ms}$ an $n = 2$ signal is located near $\psi_N = 0.44$ but fails the check on $n\,q$, so we identify no resonant structure at S50 at that time, only shortly afterwards; the band still enters $s_{\mathrm{MHD}}$, which is therefore low at S50 (Section~\ref{sec:analysis-surfaces}).

\subsection{Selecting candidate surfaces}
\label{sec:candidate-surfaces}

\begin{figure}[tbp]
  \centering
  \includegraphics{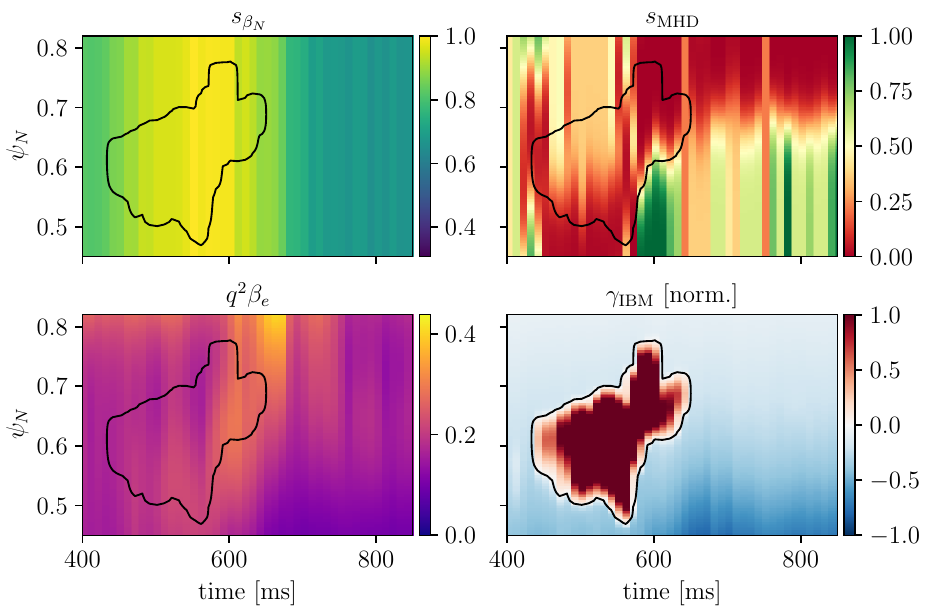}
  \caption{The four ingredients of the score $S_{\mathrm{GK}}$ of~(\ref{eq:gk-score}) over $(\psi_N,\,t)$: the pressure factor $s_{\beta_N}$ and the magnetic activity factor $s_{\mathrm{MHD}}$ (top row); the drive factor $q^2\beta_e$ and the ideal-ballooning eigenvalue $\gamma_{\mathrm{IBM}}$ in $(c_s/a)^2$ (bottom row). The black curve is $\gamma_{\mathrm{IBM}} = 0$.}
  \label{fig:criteria-maps}
\end{figure}

Figures~\ref{fig:criteria-maps} and~\ref{fig:score-pair} show the factors and the score of~(\ref{eq:gk-score}) respectively as functions of $(\psi_N,t)$. Two features are evident. Firstly, the gate removes an IBM-unstable pocket at peak $\beta_N$ just outboard of $\psi_N = 0.50$, and $q^2\beta_e$ peaks just inside its stable side. The ideal eigenvalue in the pocket is a few per cent of its values well inside the unstable region of the $(\hat{s},\,\alpha_{\mathrm{MHD}})$ plane (Section~\ref{sec:ibm}), so the pocket is close to marginal rather than strongly unstable. Whether the reconstructed equilibrium is reliable where it is ideal-unstable is a question for the equilibrium reconstruction that we do not address here. Secondly, the score is concentrated in a band along the stable side of the IBM boundary, where the drive is strongest and the local model is expected to hold. The lower panels of Figure~\ref{fig:score-pair} show the total \texttt{TRANSP} electron and ion heat fluxes, $Q_e$ and $Q_i$, which are the reference fluxes against which we compare the nonlinear simulations in Section~\ref{sec:nonlinear}. 

Cyan stars mark the locations of the two surfaces selected by this analysis in Figure~\ref{fig:score-pair}. S50 ($t = 531\,\mathrm{ms}$, $\psi_N = 0.50$) [close to peak $\beta_N,$ also examined in \cite{patel2026etg}] and S60 ($t = 591\,\mathrm{ms}$, $\psi_N = 0.60$) [at the maximum IBM stable $q^2 \beta_e$].

\subsubsection{Reconciling our analysis with \texttt{TRANSP} modelling}
\label{sec:transp-reconciliation}

\begin{figure}[tbp]
  \centering
  \includegraphics{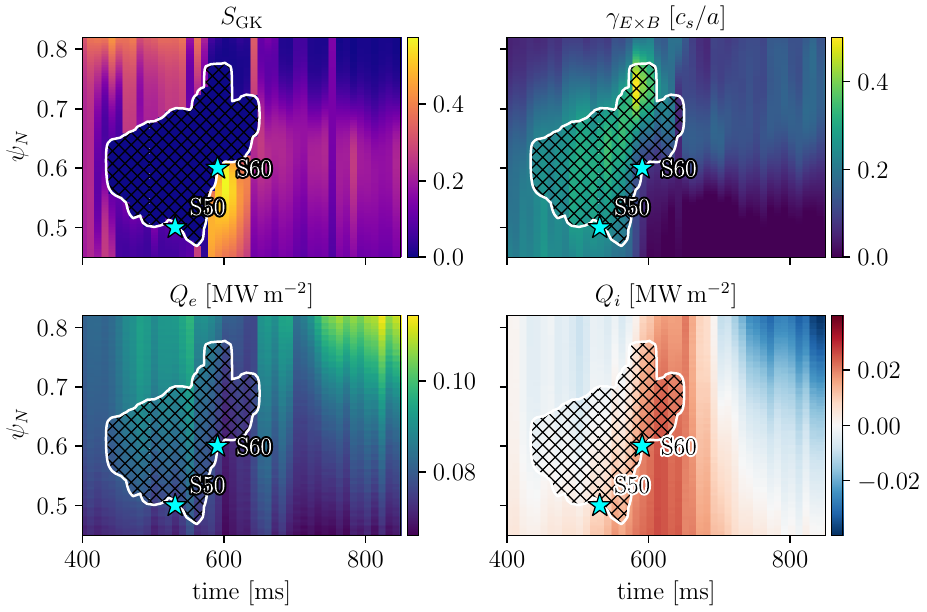}
  \caption{The score $S_{\mathrm{GK}}$ of~(\ref{eq:gk-score}) and the $E\times B$ shearing rate (top row), and the total \texttt{TRANSP} electron and ion heat fluxes per unit area, $Q_e$ and $Q_i$ in $\mathrm{MW\,m^{-2}}$ (bottom row), over $(\psi_N,\,t)$. The fluxes are conducted plus convected, including the neoclassical part, and averaged over $\pm20\,\mathrm{ms}$. White curve, $\gamma_{\mathrm{IBM}} = 0$; cross-hatching, the IBM-unstable pocket. Cyan stars mark S50 and S60.}
  \label{fig:score-pair}
\end{figure}

We compare the heat flux of the nonlinear simulations, summed over the kinetic species, with the total \texttt{TRANSP} power-balance heat flux at the same surface and time, $Q_{\mathrm{TRANSP}} \equiv Q_e + Q_i$, averaged over $\pm20\,\mathrm{ms}$ (Figure~\ref{fig:score-pair}, bottom row): $0.082$ and $0.094\,\mathrm{MW\,m^{-2}}$ at S50 and S60. We do not subtract the neoclassical part, which is uncertain; subtracting it would lower $Q_{\mathrm{TRANSP}}$ by about a tenth and change no result. We regard this comparison as indicative rather than definitive. \texttt{TRANSP} is an interpretive power-balance analysis: it knows only the equilibrium, the profiles and the sources, and is only as good as the experimental constraints on them. The total flux may also include transport driven by the $2/1$ neoclassical tearing mode or by other low-$n$ modes, and the beam heating depends on a prescribed fast-ion diffusivity.

Importantly, the surfaces are not chosen to show that ion-scale turbulence carries the experimental heat flux. Instead, we aim to assess the proximity of \#48657 to the non-zonal transition;  this humbler objective should be relatively robust to uncertainties associated with assumptions made in \texttt{TRANSP}. A surface that lies just below the transition, or inside the no-go zone and seemingly held there by flow shear, ought to carry transport fluxes that are balanced with sources, whilst lying close to a regime of much larger fluxes. In this work, $Q_{\mathrm{TRANSP}}$ thus serves only as a reference for the level of the heat flux.

\subsection{The two analysis surfaces}
\label{sec:analysis-surfaces}

We analyse two surfaces. S60 ($t = 591\,\mathrm{ms}$, $\psi_N = 0.60$) lies in the band of high score along the stable side of the IBM boundary (Section~\ref{sec:candidate-surfaces}), at the largest IBM-stable $q^2\beta_e$ inside $\psi_N = 0.8$ at that time, where the drive is strongest. The nearest coherent structure, on the $q = 3$ surface, lies $0.15$ outboard in $\psi_N$ and lowers its score by about a third. S50 ($t = 531\,\mathrm{ms}$, $\psi_N = 0.50$) is the surface of the companion study~\cite{patel2026etg}, close to peak $\beta_N$, and we include it for comparison rather than for its score. No band passes the check on $n\,q$ near S50 at that time, but the $n = 2$ band placed near $\psi_N = 0.44$ still enters $s_{\mathrm{MHD}}$, so that $s_{\mathrm{MHD}} \approx 0.04$ at S50; $n = 2$ activity that passes the check appears at its radius from $544\,\mathrm{ms}$. S50 lies just inside the $q = 2$ surface ($|q| = 1.94$), and S60 ($|q| = 2.22$) outboard of it; on that surface the $2/1$ neoclassical tearing mode that drives the rotation collapse forms later, at $t \approx 600\,\mathrm{ms}$~\cite{harrison2026, patel2026etg}. We do not study a surface after the rotation collapse, since the outer core then lies in the band of coherent low-$n$ activity between $\psi_N = 0.75$ and $0.9$.

\section{Local equilibrium and linear stability of the analysis surfaces}
\label{sec:equilibrium-linear}

In this section, we determine how close S50 and S60 are to the threshold for ideal and kinetic ballooning modes, below which the non-zonal transition occurs in STEP~\cite{kennedy2026}. For the majority of this section, except where explicitly mentioned, we neglect the effects of the equilibrium shear rotation. In such a system, local ideal stability is set by the magnetic shear and the normalised pressure gradient,
\begin{equation}
  \hat{s} \equiv \frac{r}{q}\,\frac{\mathrm{d}q}{\mathrm{d}r},
  \qquad
  \alpha_{\mathrm{MHD}} \equiv -q^{2}\,\frac{R}{a}\,\beta^{\prime},
  \label{eq:shat_alpha_def}
\end{equation}
where $r$ and $R$ are the minor and major radii and $\beta^{\prime}$ is the normalised pressure gradient of~(\ref{eq:app_betaprime}). In this plane, the IBM boundary separates the stable and unstable regions.

\subsection{Ideal-ballooning stability}
\label{sec:ibm}

We first examine ideal ballooning stability of MAST-U which entered the scoring of Section~\ref{sec:candidate-selection}, with the IBM boundaries calculated using \texttt{pyrokinetics}~\cite{pyrokinetics2024}. An IBM-unstable pocket develops close to peak performance in the discharge (Figure~\ref{fig:criteria-maps}, bottom right panel). At $t = 531\,\mathrm{ms}$ it spans $\psi_N \approx 0.51$ to $0.69$, with S50 sitting on its stable inner edge. The S50 surface then enters the pocket and leaves it again, and by $t = 591\,\mathrm{ms}$ the pocket has moved outward, to $\psi_N \gtrsim 0.61$, leaving S60 close to the IBM boundary. This locates in radius the IBM-marginal part of the $\beta_e$ profile reported in~\cite{harrison2026} at $600\,\mathrm{ms}$. The Mirnov coils cannot detect such high $n$ modes.\footnote{The ideal-ballooning solver returns the local $n \to \infty$ limit; close to this ideal limit, the KBM can dominate the turbulence~\cite{Pueschel2008}.}

\begin{figure}[tbp]
  \centering
  \includegraphics[width=\textwidth]{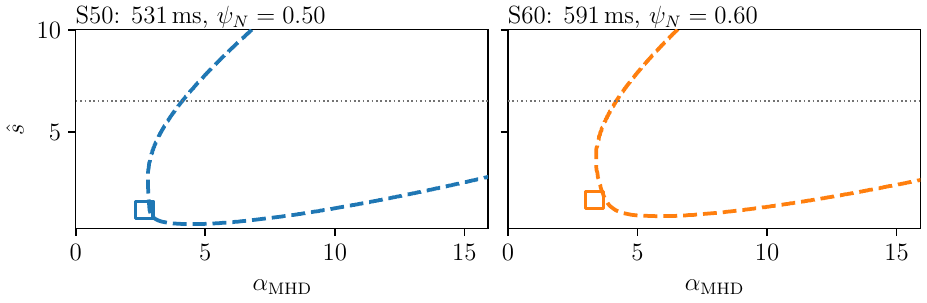}
  \caption{Ideal-ballooning stability of the two analysis surfaces over $\hat{s} \in [0.25,\,10]$ and $\alpha_{\mathrm{MHD}} \in [0,\,6\alpha_{\mathrm{MHD}}^{\mathrm{op}}]$, with $\alpha_{\mathrm{MHD}}^{\mathrm{op}}$ the experimental value: S50 (\textcolor{tabblue}{blue}, left-hand side) and S60 (\textcolor{taborange}{orange}, right-hand side). Dashed contours, the ideal-ballooning boundary, computed on a $384\times384$ grid; open square, the experimental operating point; dotted horizontal, $\hat{s} = 6.5$, the ceiling to which the boundary is computed in the scans of Section~\ref{sec:salpha}.}
  \label{fig:ibm-wide}
\end{figure}

Static stability maps over a much larger region of the plane are shown in Figure~\ref{fig:ibm-wide}. Both equilibria are IBM-stable but sit close to their respective boundaries, where EM turbulence is expected to be active; Table~\ref{tab:ibm-neighbours} gives their local $\hat{s}$, $\alpha_{\mathrm{MHD}}$ and ideal-ballooning eigenvalues. 

\begin{table}[tbp]
  \centering
  \caption{Local $\hat{s}$, $\alpha_{\mathrm{MHD}}$, ideal-ballooning eigenvalue $\gamma_{\mathrm{IBM}}$ in $(c_s/a)^2$, negative where stable, and selection score $S_{\mathrm{GK}}$ of~(\ref{eq:gk-score}) at the two analysis surfaces.}
  \label{tab:ibm-neighbours}
  \begin{tabular*}{\textwidth}{@{\extracolsep{\fill}}ccccccl@{}}
\toprule
$t$\,[ms] & $\psi_N$ & $\hat{s}$ & $\alpha_{\mathrm{MHD}}$ & $\gamma_{\mathrm{IBM}}$ & $S_{\mathrm{GK}}$ & verdict \\
\midrule
531 & 0.50 & 1.18 & 2.65 & $-4.1\times 10^{-4}$ & 0.095 & IBM-stable \\
591 & 0.60 & 1.66 & 3.34 & $-3.2\times 10^{-4}$ & 0.393 & IBM-stable \\
\bottomrule
\end{tabular*}

\end{table}

\subsection{Linear microstability} 
\label{sec:linear}
The two local equilibria are IBM stable but sit close to the boundary and might therefore be expected to host long-wavelength EM GK instabilities.

\subsubsection{Numerical setup}
\label{sec:linear-setup}

We perform local linear electromagnetic simulations with \texttt{GENE}~\cite{Jenko2000, gene}, using inputs generated with \texttt{pyrokinetics}~\cite{pyrokinetics2024}. Details of the local equilibrium parameters are given in Table~\ref{tab:gk_setup} and details of the numerical discretisation grids\footnote{One non-standard choice of numerical scheme is made in this work and in previous studies of EM instabilities (e.g.,~\cite{kennedy2023a}). Namely, to avoid numerical instability we replace the conservative Arakawa bracket (which conserves the gyrokinetic free energy) in the parallel operator with centred differences (which do not). When the Arakawa bracket is used, the fastest-growing unstable mode at long-wavelengths is a grid-scale numerical instability. At shorter wavelengths, the two schemes agree to within about three percent. In \ref{app:arakawa}, we construct a conservative corrected operator and show that the physics of the system is unchanged by this choice beyond allowing removal of the spurious numerical instability at more modest resolution due to the additional damping centred differences introduces. The $(\hat{s},\,\alpha_{\mathrm{MHD}})$ scans of Section~\ref{sec:salpha} are the one exception: they use the Arakawa bracket, at $k_y\rho_s = 0.3$, where the two schemes agree to within about one per cent at the operating point (\ref{app:arakawa}).} are given in Table~\ref{tab:numerics}. Wavenumbers are normalised to the sound Larmor radius $\rho_s$, with $k_y$ the binormal wavenumber. We retain all three fields in our simulations, since the omission of $\dbp$ at these values of $\beta_e$ distorts ion-scale growth rates~\cite{kennedy2024} (and is also known to impact strongly on ETG growth rates \cite{Roach2005,patel2026etg}). Collisions use \texttt{GENE}'s linearised Landau operator with finite-Larmor-radius corrections; between $k_y\rho_s \approx 0.1$ and the peak of the growth-rate spectrum, the Sugama operator changes the growth rate by less than a tenth; at longer wavelengths, where the growth rates are small, the two differ by more (\ref{app:collisions}).

\subsubsection{Species treatment and pressure gradient}
\label{sec:linear-species}

\begin{figure}[tbp]
  \centering
  \includegraphics{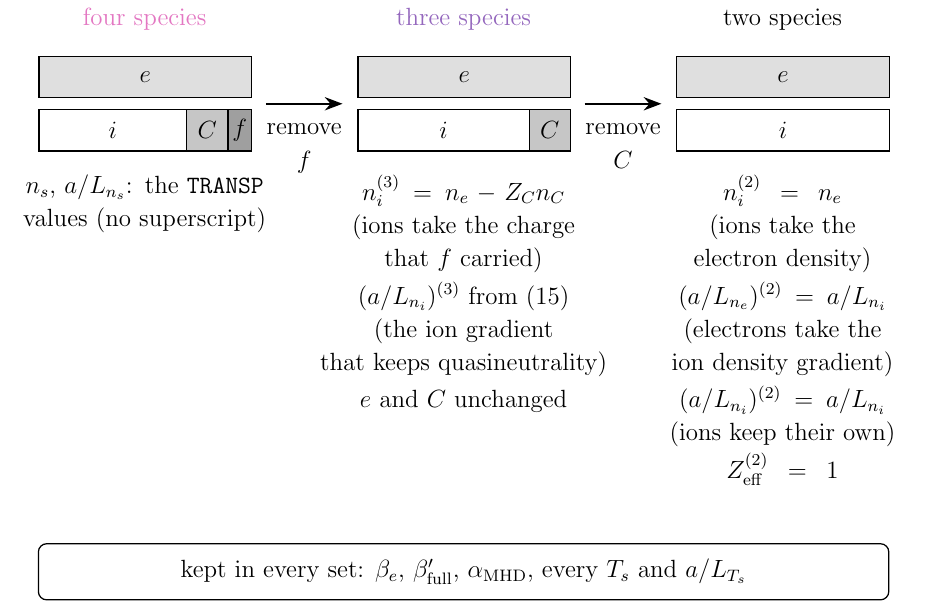}
  \caption{Schematic of the four, three and two kinetic-species sets (\textcolor{tabpink}{pink}, \textcolor{tabpurple}{purple} and black). Top bar, the electron density; bottom bar, the ion charge density $Z_s n_s$ of thermal deuterium ($i$), carbon ($C$) and fast beam deuterium ($f$), not to scale. Below each set, its rules for the densities and their gradients, as in~(\ref{eq:species_two}) and~(\ref{eq:species_three}) of Section~\ref{sec:species-three-four}: a superscript $(2)$ or $(3)$ marks the reduced set's value, set equal to the four-species value on the right, and the words in brackets say which species takes what from which. Box, the quantities common to all three sets.}
  \label{fig:species-cartoon}
\end{figure}

In the majority of this work we evolve only two kinetic species, electrons and thermal deuterium. At each surface, \texttt{TRANSP} provides four species: electrons ($e$), thermal deuterium ($i$), carbon ($C$, with charge number $Z_C = 6$), and fast beam deuterium ($f$); Table~\ref{tab:species_inputs} of \ref{app:species-table} gives the local parameters for all species. Each species $s$ has a charge number $Z_s$, a density $n_s$, a temperature $T_s$, and inverse gradient lengths $a/L_{n_s} = -a\,\mathrm{d}\ln n_s/\mathrm{d}r$ and $a/L_{T_s}$. The equilibrium and its radial gradient are quasineutral,
\begin{equation}
  \sum_s Z_s n_s = 0,
  \qquad
  \sum_s Z_s n_s\,\frac{a}{L_{n_s}} = 0,
  \label{eq:species_qn}
\end{equation}
where $Z_e = -1$. Carbon carries about a tenth of the ion charge and the fast ions about one per cent; the fast ions are hot, with $T_f \gg T_e$, and strongly peaked. To reduce this to two species, following~\cite{Bokshi2025}, we keep the electrons (apart from the density gradient) as they are, give the thermal deuterium the electron density, and give both species the \texttt{TRANSP} deuterium density gradient,
\begin{equation}
  n_i^{(2)} = n_e,
  \qquad
  \left(\frac{a}{L_{n_e}}\right)^{(2)} = \left(\frac{a}{L_{n_i}}\right)^{(2)} = \frac{a}{L_{n_i}},
  \qquad
  Z_{\mathrm{eff}}^{(2)} = 1,
  \label{eq:species_two}
\end{equation}
where a superscript $(2)$ marks the value in the reduced set and a quantity without one is the four-species \texttt{TRANSP} value. Every $Z_s$, $T_s$ and $a/L_{T_s}$ is unchanged, so that~(\ref{eq:species_qn}) holds with a single, singly charged ion, and $\beta_e$ is unchanged. Three things are lost during this merge. The kinetic response of carbon and of the fast ions is discarded, whilst their pressure stays in the equilibrium through $\beta'$ (see discussion below). $Z_{\mathrm{eff}}$ is not preserved, it falls from $Z_\mathrm{eff} = 1.5$ to unity.\footnote{Merging the ion species in this way changes the physics in two ways beyond the loss of the carbon response: the impurity shielding of the electrostatic potential (Section~\ref{sec:species-nonlinear}) is lost, and the electrons are less collisional, because the electron--ion collision frequency scales with $Z_{\mathrm{eff}}$ and the collisions with carbon are gone.} The electron density gradient takes the deuterium value, which is below the \texttt{TRANSP} $a/L_{n_e}$ (Table~\ref{tab:species_inputs}). Figure~\ref{fig:species-cartoon} illustrates the reduction. Later, in Section~\ref{sec:species}, we also show the results of calculations with three kinetic species (retaining the carbon impurity) and four kinetic species (retaining the carbon impurity and the fast beam deuterium, as a hot Maxwellian).

Instabilities such as the KBM are driven by the pressure gradient, which also enters the magnetic drifts and the field-line bending term in the GK equation via
\begin{equation}
  \beta' \equiv \frac{2\mu_0 a}{B_0^2}\,\frac{\mathrm{d}p}{\mathrm{d}r},
  \label{eq:app_betaprime}
\end{equation}
where $p$ is the equilibrium pressure, $a$ the minor radius and $B_0$ the reference magnetic field; $\alpha_{\mathrm{MHD}}$ follows from $\beta'$ by~(\ref{eq:shat_alpha_def}). Once one or more species are removed, $p$ has two straightforward possible choices,
\begin{equation}
  \beta'_{\mathrm{full}} = \frac{2\mu_0 a}{B_0^2}\,\frac{\mathrm{d}}{\mathrm{d}r}
  \left(\sum_{s} n_s T_s + p_{\mathrm{rot}}\right),
  \qquad
  \beta'_{\mathrm{cons}} = \frac{2\mu_0 a}{B_0^2}\,\frac{\mathrm{d}}{\mathrm{d}r}
  \big(n_e T_e + n_i T_i\big)\Big|_{n_i = n_e}.
  \label{eq:app_two_sums}
\end{equation}
Here, the full-plasma $\beta'_{\mathrm{full}}$ sums the pressures $n_s T_s$ of the four \texttt{TRANSP} species (Table~\ref{tab:species_inputs}), electrons, thermal deuterium, carbon, and fast beam deuterium, \emph{and the rotation pressure} $p_{\mathrm{rot}}$, whilst the species-consistent $\beta'_{\mathrm{cons}}$ keeps only the kinetic pressure from the two evolved species. The rotation pressure entering (\ref{eq:app_two_sums}) represents the centrifugal force on the ions, which acts alongside the standard pressure gradient in the equilibrium force balance,
\begin{equation}
  \nabla p_{\mathrm{kin}} = \mathbf{J}\times\mathbf{B} + \rho_m\,\omega_0^2\,\mathbf{R},
  \label{eq:rot_force_balance}
\end{equation}
where $p_{\mathrm{kin}}$ is the thermal and fast-ion pressure, $\mathbf{J}$ the current density, $\mathbf{B}$ the magnetic field, $\rho_m$ the ion mass density and $\mathbf{R}$ the position vector from the symmetry axis, perpendicular to it. \texttt{TRANSP} adds this term, of the order of the rotational kinetic energy density $\tfrac{1}{2}\rho_m\omega_0^2|\mathbf{R}|^2$, to the thermal and fast-ion pressures in its equilibrium solver. At S50 and S60 it carries less than a tenth of the pressure gradient.

In this work, as in~\cite{patel2026etg}, we choose to keep $\beta$, $\beta^{\prime}$ and $\alpha_{\mathrm{MHD}}$ at their four-species values such that $\beta^\prime \equiv \beta^\prime_\mathrm{full}$, which differs from $\beta'_{\mathrm{cons}}$; the two could be made equal by adjusting the temperature gradients of the evolved species, which we do not do. Whenever $\beta_e$ is increased (e.g., in scans), $\beta^\prime$ is rescaled by the same factor, e.g., $\beta_e \rightarrow m_\beta\beta_e \implies \beta^\prime \rightarrow m_\beta\beta^\prime$ for some $m_\beta \in \mathbb{R}.$ This scales every other contribution to $\beta'$, including the rotation pressure, by the same factor as the electron one; the normalisation to $T_e$ and $n_e$ is unchanged, so the larger absolute pressure enters only through $\beta_e$ and $\beta'$.

The motivation for using the MHD equilibrium value of $\beta^\prime$ is that the fast-ion orbits in MAST-U span a sizeable fraction of the minor radius, formally outside the local ordering, whilst force balance keeps their pressure in $\beta'$ (i.e., the fast ions are not obviously well-described by local gyrokinetics but still contribute to the equilibrium). With the species-consistent $\beta'_{\mathrm{cons}}$ instead, the ballooning peak at S60 is lowered and the microtearing peak at S50 raised somewhat, but the character of neither spectrum changes (not shown): the contribution of the fast ions (and of the rotation) to $\beta'$, about a fifth of it, does not seem to be important for the linear spectra of this plasma. This is consistent with the companion study~\cite{patel2026etg}, which finds that removing the fast ions in this way make little difference to electron-scale spectra.

\begin{table}[tbp]
  \caption{Local gyrokinetic parameters of the two analysis surfaces as used in \texttt{GENE}; the four-species \texttt{TRANSP} parameters from which they are set are in Table~\ref{tab:species_inputs}. Here $R$ is the major radius of the centre of the flux surface, $\kappa$ and $\delta$, the elongation and triangularity, are the leading MXH shape moments, $\beta^\prime$ is the full four-species value including rotation as defined in~(\ref{eq:app_two_sums}), $\rho_s = \sqrt{T_e m_i}/(e B_0)$ with $c_s = \sqrt{T_e/m_i}$, $\rho_* = \rho_s/a$, 
  $\nu_{ei}$ is the electron-ion collision frequency, equal to the electron-electron one because the two-species model has a single, singly charged ion species of the electron density; $n_i/n_e$, with $n_i$ the thermal deuterium density, is set by the two-species reduction~(\ref{eq:species_two}); and $a$ is the minor radius. $a/L_{T_e}$ and $a/L_{T_i}$ are the normalised electron and ion temperature gradients, and $a/L_{n_e}$ is the bulk-ion density gradient that the reduction gives both species, with $a/L_X = -a\,\mathrm{d}\ln X/\mathrm{d}r$. $\gamma_{E\times B}$ is the measured shearing rate, zero in the static simulations and applied in the rotating linear simulations and the nonlinear flow-shear scan. $\alpha_{\mathrm{MHD}}$ is given in Table~\ref{tab:ibm-neighbours}.}
  \label{tab:gk_setup}
  \centering
  \begin{tabular*}{\textwidth}{@{\extracolsep{\fill}}lcclcc@{}}
\toprule
$t$\,[ms] & 531 & 591 & $t$\,[ms] & 531 & 591 \\
\midrule
$\psi_N$ & 0.50 & 0.60 & $\rho_* = \rho_s/a$ & 0.0143 & 0.0144 \\
$r/a$ & 0.62 & 0.68 & $T_e$\,[keV] & 0.76 & 0.77 \\
$R/a$ & 1.73 & 1.72 & $n_e$\,[$10^{19}\,\mathrm{m^{-3}}$] & 4.93 & 4.88 \\
$\kappa$ & 1.74 & 1.75 & $T_i/T_e$ & 1.34 & 1.17 \\
$\delta$ & 0.13 & 0.13 & $n_i/n_e$ & 1.00 & 1.00 \\
$q$ & $-1.94$ & $-2.22$ & $a/L_{n_e}$ & 0.68 & 1.16 \\
$\hat{s}$ & 1.18 & 1.66 & $a/L_{T_e}$ & 1.37 & 1.77 \\
$\beta_e$ & 0.0580 & 0.0567 & $a/L_{T_i}$ & 1.66 & 0.76 \\
$\beta^\prime$ & $-0.41$ & $-0.39$ & $(a/c_s)\nu_{ei}$ & 0.384 & 0.372 \\
$q^2\beta_e$ & 0.22 & 0.28 & $(a/c_s)\gamma_{E\times B}$ & 0.251 & 0.096 \\
$\rho_s$\,[mm] & 7.81 & 7.76 &  &  &  \\
\bottomrule
\end{tabular*}

\end{table}

\begin{table}[tbp]
  \caption{Nominal resolution and box parameters of the linear (L) and nonlinear (NL) local (LOC) simulations. Here $n_{k_x}$ and $n_{k_y}$ are the numbers of radial and binormal Fourier modes, $n_z$ the number of points along the field line, and $n_{v_\parallel}$ and $n_\mu$ the velocity-space resolutions; $L_x$ and $L_y$ are the radial and binormal box sizes in units of $\rho_s$, $k_x^{\mathrm{min}}\rho_s$ is the smallest finite radial wavenumber, and $n_0^{\mathrm{min}}$ and $n_0^{\mathrm{max}}$ the smallest and largest toroidal mode numbers evolved. The linear simulations evolve one binormal mode at a time, spanning $k_y\rho_s \approx 0.021$ to $11$ across the two surfaces, with $k_y\rho_s \in \{0.1,\,0.2,\,0.3\}$ in the $(\hat{s},\,\alpha_{\mathrm{MHD}})$ scans. $N_\theta$ is the number of poloidal turns of the mode centred at $\theta = 0$ at $k_y\rho_s = 0.2$, with $\theta$ the extended ballooning angle; the linear simulations resolve thirteen turns. The nonlinear simulations have one row per surface. Every simulation outside the $(\hat{s},\,\alpha_{\mathrm{MHD}})$ scans uses the centred parallel differences with hyperdiffusion of~\ref{app:arakawa}; the $(\hat{s},\,\alpha_{\mathrm{MHD}})$ scans use the Arakawa bracket (Section~\ref{sec:linear-setup}).}
  \label{tab:numerics}
  \centering
  \renewcommand{\arraystretch}{1.5}
  \small
  \begin{tabular*}{\textwidth}{@{\extracolsep{\fill}}lccccccccccc@{}}
\toprule
\textbf{Simulation} & \( n_{k_x} \) & \(n_{k_y} \) & \( n_{z} \) & \( n_{v_\parallel} \) & \( n_{\mu} \) & \( L_{x} \) & \( k_{x}^{\mathrm{min}} \) & \( L_y \) & \( n_{0}^{\mathrm{min}} \) & \( n_{0}^{\mathrm{max}} \) & \( N_{\theta} \) \\ \midrule
L (LOC) & 13 & 1 & 70 & 56 & 28 & -- & -- & -- & 1 & 500 & 13 \\
L (LOC), $(\hat{s},\,\alpha_{\mathrm{MHD}})$ & 13 & 1 & 70 & 56 & 28 & -- & -- & -- & -- & -- & 13 \\
NL (LOC), S50 & 256 & 24 & 32 & 24 & 14 & 165 & 0.038 & 157 & 2 & 46 & 3 \\
NL (LOC), S60 & 256 & 24 & 32 & 24 & 14 & 108 & 0.058 & 157 & 2 & 46 & 3 \\
\bottomrule
\end{tabular*}

\end{table}

\subsubsection{Ion-scale spectra}
\label{sec:linear-spectra}

\begin{figure}[tbp]
  \centering
  \includegraphics{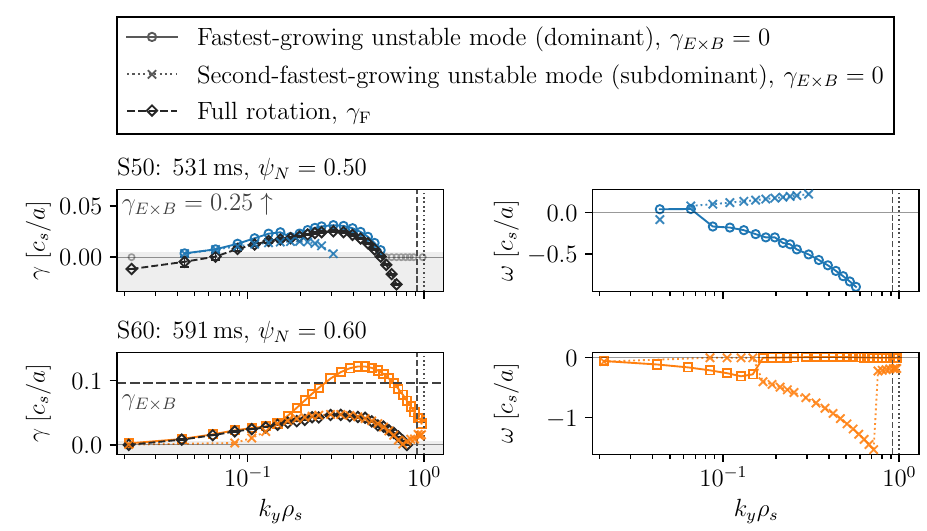}
  \caption{Growth rate (left-hand side) and real frequency, positive in the ion direction (right-hand side), of the two-species modes at the full-plasma $\beta'$ against $k_y\rho_s$, one row per surface (S50 \textcolor{tabblue}{blue} circles, S60 \textcolor{taborange}{orange} squares; every marker is open). Solid curves, the fastest-growing unstable mode (dominant) at $\gamma_{E\times B} = 0$: coloured markers, converged growing modes; small grey markers, damped or unconverged modes; the curve joins the coloured markers only. Dotted curves and crosses, the second-fastest-growing unstable mode (subdominant) at $\gamma_{E\times B} = 0$ of the same simulations by dynamic mode decomposition~\cite{Schmid2010,Dudkovskaia2025}, drawn where it is unstable and broken where it changes direction. Black diamonds, the Floquet-averaged growth rate under full rotation (\ref{app:floquet}), with bars spanning its spread over whole periods. Grey band, $|\gamma| < 5\times10^{-3}\,c_s/a$; horizontal dashed line, $\gamma_{E\times B}$ of the surface, quoted at the top left where it lies above the panel; dashed vertical, the maximum $k_y \rho_s$ in the nonlinear simulation at $k_y\rho_s = 0.92$; dotted vertical, the start of the electron-scale scan of~\cite{patel2026etg}.}
  \label{fig:spectra}
\end{figure}

Figure~\ref{fig:spectra} shows the growth rate $\gamma$ (left-hand side) and real frequency $\omega$ (right-hand side) of the fastest-growing unstable mode (dominant) against $k_y\rho_s$, without equilibrium flow shear. The ion-scale, at $\gamma_E=0$, is unstable on both surfaces. At S50, the fastest-growing unstable mode is a microtearing mode (mode structures are shown in Figure~\ref{fig:mode_structure}), as reported in ~\cite{patel2026etg}, propagating in the electron diamagnetic direction, and growing an order of magnitude more slowly than the shearing rate $\gamma_{E\times B}$. At S60, the fastest-growing unstable mode is instead a ballooning-like mode with a peak growth rate several times larger, and barely propagating ($|\omega| < 0.01\,c_s/a$). For each surface, the second-fastest-growing unstable mode (subdominant) can be recovered using dynamic mode decomposition~\cite{Schmid2010,Dudkovskaia2025} (DMD). On both surfaces a second mode is unstable at both surfaces, propagating in the opposite direction. Similar to STEP and to other conceptual high-$\beta$ STs (e.g., \cite{Dudkovskaia2026}), and to low-collisionality NSTX plasmas~\cite{singh2026}, a mixture of hKBMs and MTMs are unstable. Electron-scale instabilities are discussed in~\cite{patel2026etg}.

With equilibrium flow shear at its measured value (full rotation), we use the Floquet-averaged growth rate $\gamma_{\mathrm{F}}$~\cite{maeyama2025floquet} of~\ref{app:floquet} (black diamonds in Figure~\ref{fig:spectra}) to assess how susceptible the linear instabilities are to stabilisation by equilibrium flow shear. Flow shear reduces the peak growth rate by about a fifth at S50, but by about three fifths at S60, whose shearing rate is the smaller, so that neither $\gamma_{E\times B}/\gamma$ nor the shearing rate alone predicts the reduction. On both surfaces, the Floquet-averaged growth rates match the growth rates of the MTM (which is the dominant instability on S50 and subdominant on S60). Consistent with previous work, the ballooning-like instabilities appear to be stabilised by shear flows~\cite{kennedy2023a} whilst the MTMs are more resilient~\cite{patel2025exb}.

\subsubsection{Mode structure}
\label{sec:linear-mode-structure}

\begin{figure}[tbp]
  \centering
  \includegraphics{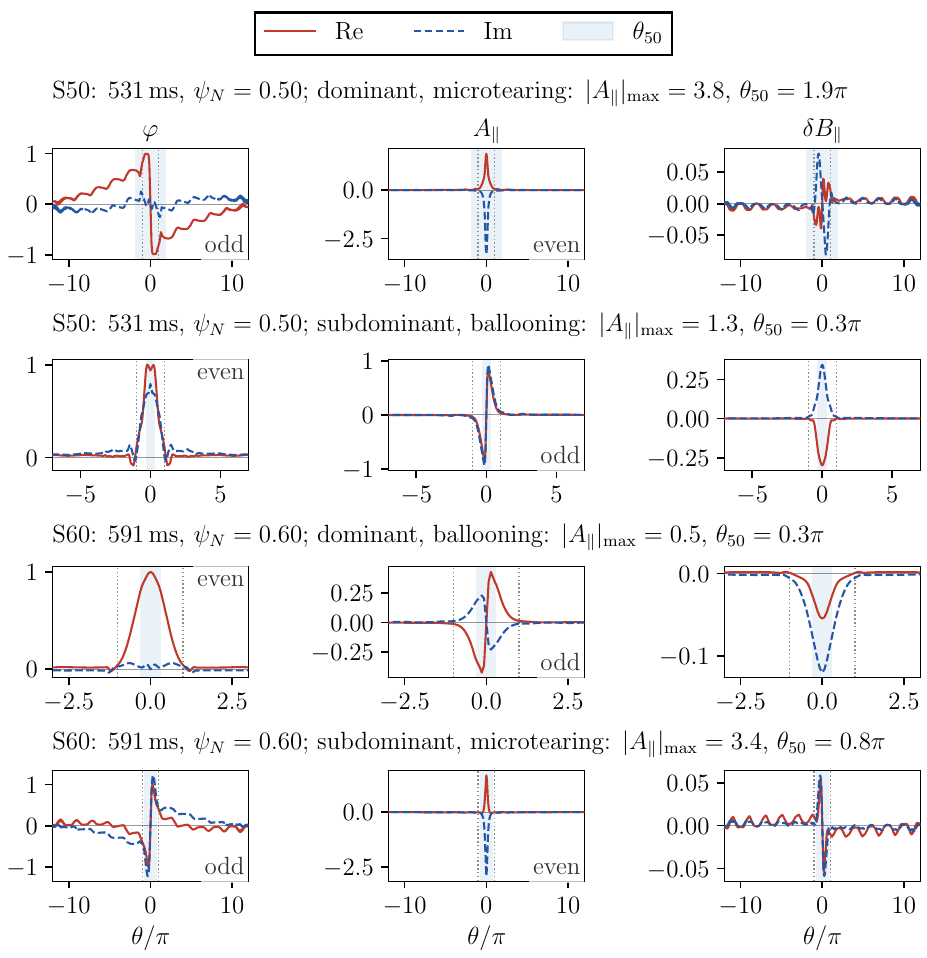}
  \caption{Parallel structure of the fastest-growing unstable mode (dominant) and the second-fastest-growing unstable mode (subdominant) at $k_y\rho_s \approx 0.2$ against the extended ballooning angle: $\varphi$ (left-hand side), $A_\parallel$ (middle) and $\dbp$ (right-hand side), each normalised to its own $\max[\mathrm{Re}\,\varphi] = 1$; rows, from the top, the dominant mode at S50, the second-fastest-growing unstable mode (subdominant) at S50 from dynamic mode decomposition~\cite{Schmid2010,Dudkovskaia2025}, and the same two at S60. Solid red, real part; dashed blue, imaginary part; shading, the half-width $\theta_{50}$ holding half the power of $\varphi$; dotted verticals, the central ballooning well. Each panel is labelled with the parity of its field about $\theta = 0$, and each row with its surface, mode character, $|A_\parallel|_{\max}$ and $\theta_{50}$. Each row is drawn out to the angle holding $99$ per cent of the power of $\varphi$.}
  \label{fig:mode_structure}
\end{figure}

A mode with $\varphi$ even and $A_\parallel$ odd about $\theta = 0$, where $\theta$ is the extended ballooning angle, has ballooning parity, and the reverse is tearing parity. Figure~\ref{fig:mode_structure} shows $\varphi$ (left-hand side), $A_\parallel$ (middle) and $\dbp$ (right-hand side) of both modes at $k_y\rho_s \approx 0.2$. At S50 the fastest-growing unstable mode (dominant) has tearing parity, with $\varphi$ extended along the field line and $|A_\parallel|$ several times larger than $|\varphi|$ consistent with MTM signatures~\cite{Patel2022}. At S60 the dominant instability has ballooning parity, with $\varphi$ localised to the central ballooning well and $|A_\parallel|$ half as large, as for a hKBM. The $\dbp$ eigenmode is small on both surfaces, and the second-fastest-growing unstable mode (subdominant) has the \emph{opposite} parity. The main difference between the surfaces is that the ballooning mode is far more unstable at S60 and dominates there when $\gamma_{E\times B} = 0$. This is not obvious from any one local parameter: S60 has the larger $q$, $\hat{s}$, $\alpha_{\mathrm{MHD}}$ and $a/L_{n_e}$ but the smaller $a/L_{T_i}$ (Tables~\ref{tab:ibm-neighbours} and~\ref{tab:gk_setup}), and both surfaces lie about a tenth below the ideal boundary in $\alpha_{\mathrm{MHD}}$. The $(\hat{s},\,\alpha_{\mathrm{MHD}})$ scans of Section~\ref{sec:salpha} show that at $k_y\rho_s = 0.3$ the ballooning mode takes over from the microtearing mode below the operating $\alpha_{\mathrm{MHD}}$ at S60, but only at the ideal boundary at S50: S60 sits above the kinetic-ballooning onset and S50 just below it.

\subsubsection{Distance to the $\beta$ and $R/L_{T_i}$ thresholds}
\label{sec:linear-thresholds}

\begin{figure}[tbp]
  \centering
  \includegraphics{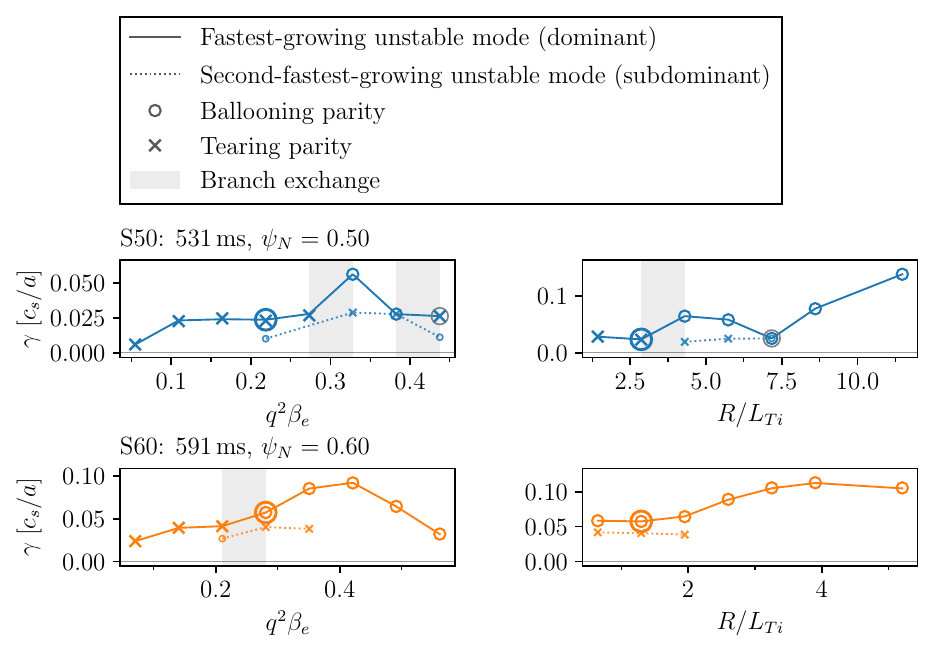}
  \caption{Growth rate at $k_y\rho_s \approx 0.2$ of the fastest-growing unstable mode (dominant, solid) and of the second-fastest-growing unstable mode (subdominant, dotted) against $q^2\beta_e$ (left-hand side) and $R/L_{T_i}$ (right-hand side); rows S50, S60. The subdominant mode is from dynamic mode decomposition~\cite{Schmid2010,Dudkovskaia2025} of the same simulations and is drawn where it is unstable. Open circles mark ballooning parity ($\varphi$ even, $A_\parallel$ odd) and crosses tearing parity about $\theta = 0$, smaller markers are used for the subdominant mode. The grey band is the interval over which the two modes exchange rank, the ringed marker the experimental operating point, and a grey ring a simulation whose \texttt{GENE} eigenvalue is not converged, drawn at its dynamic-mode-decomposition value.}
  \label{fig:param_scans}
\end{figure}

Figure~\ref{fig:param_scans} shows the growth rate at $k_y\rho_s \approx 0.2$ against $q^2\beta_e$, with $\beta'$ scaled with $\beta$ (left-hand side), and against $R/L_{T_i}$, the ion temperature gradient normalised to the major radius (right-hand side). Tearing- and ballooning-parity modes are unstable together at about half of the points, so that a change of the fastest-growing unstable mode (dominant) is simply an exchange of which instability has the largest growth rate rather than the onset of a new instability. As $q^2\beta_e$ increases, the ballooning mode becomes dominant a quarter to a half above the nominal value at S50, and just below it at S60, where it already dominates. The growth rate then more than doubles at S50 and rises by about three fifths at S60. By twice the nominal value, $\beta^{\prime}$ stabilisation~\cite{bourdelle2003} has returned S50 to a microtearing mode and halved the growth rate at S60. As $R/L_{T_i}$ increases, the ballooning mode becomes dominant at S50 at about half as much again as the nominal gradient, whereas at S60 a single ballooning mode dominates throughout, its frequency drifting through zero. 

\subsection{The $(\hat{s},\,\alpha_{\mathrm{MHD}})$ plane and the ``KBM mountain''}
\label{sec:salpha}

\begin{figure}[tbp]
  \centering
  \includegraphics[width=\textwidth]{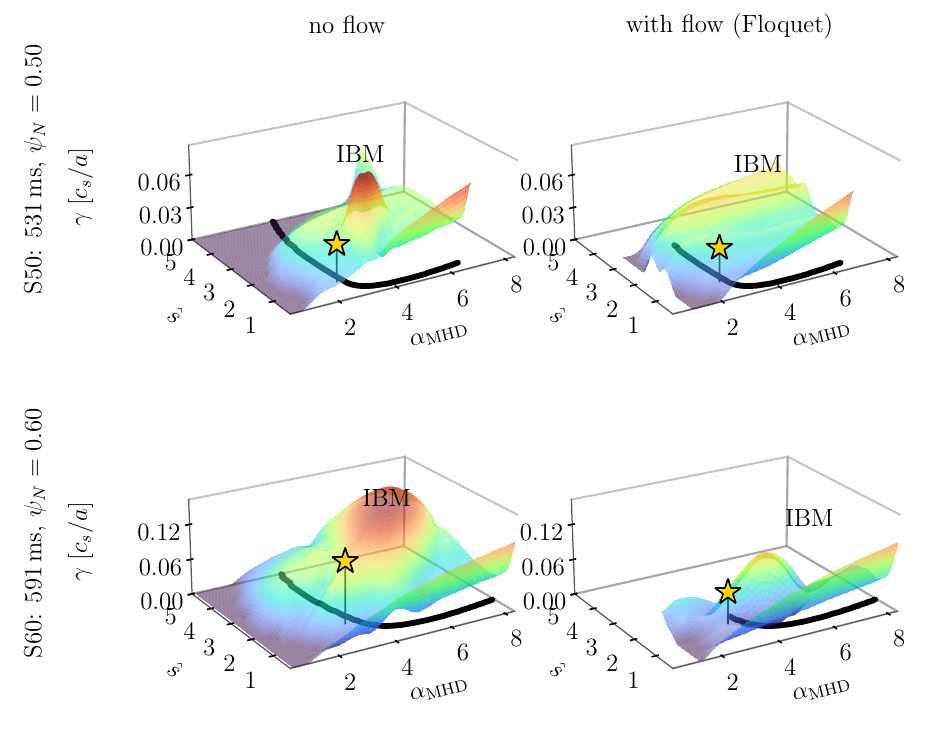}
  \caption{Ion-scale growth rate over the $(\hat{s},\,\alpha_{\mathrm{MHD}})$ plane, static (left-hand side) and Floquet-averaged under full rotation (right-hand side); rows S50, S60. Height and colour give $\gamma$ at $k_y\rho_s = 0.3$, for $\hat{s} \in [0.25,\,5]$ and a multiplier $m_\beta \in [0.125,\,2.5]$ on $(\beta,\,\beta',\,\alpha_{\mathrm{MHD}})$; damped cells are clipped at the floor, and the rotating panels omit cells holding less than one Floquet period. Every panel has the same $\hat{s}$ and $\alpha_{\mathrm{MHD}}$ ranges; where a scan has no data the base plane is empty. Star on a stem, the experimental operating point; black curve on the base plane, seen through the translucent surface, $\gamma_{\mathrm{IBM}} = 0$, with IBM marking the ideal-unstable side.}
  \label{fig:salpha_mountains}
\end{figure}

Figure~\ref{fig:salpha_mountains} shows the growth rate at $k_y\rho_s = 0.3$ over the $(\hat{s},\,\alpha_{\mathrm{MHD}})$ plane at fixed kinetic gradients, with $\beta$, $\beta'$ and $\alpha_{\mathrm{MHD}}$ scaled together by a multiplier at each $\hat{s}$, so that the pressure gradient entering the geometry and the field equations is varied independently of the temperature and density gradients that drive the modes, without flow shear (left-hand side) and Floquet-averaged with full rotation (right-hand side). This is the ``KBM mountain'' introduced by Staebler~\textit{et al.}~\cite{Staebler2016,staebler2018pop, staebler2018nf} for DIII-D (see also~\cite{mcclenaghan2019, ding2023}). On both surfaces the growth rate at $k_y\rho_s = 0.3$ rises towards the ideal boundary; S60 sits on that rise, whereas S50 sits on the flat microtearing plateau below it, and neither surface has a second-stable region within the window. The growth rate is continuous across the ideal boundary: at S50 the ballooning branch overtakes the microtearing mode there, and on both surfaces it rises inside the region, peaks, and falls again at larger $\alpha_{\mathrm{MHD}}$, where a tearing-parity mode is again the fastest. There is no sudden change in growth rate as the boundary is crossed.\footnote{Crossing the ideal boundary need not change the gyrokinetic growth rate at $k_y\rho_s = 0.3$. The ideal mode is the $n \to \infty$ limit; deep inside the ideal-unstable region it appears in local gyrokinetics as a ballooning mode unstable to the longest wavelengths, with a growth rate that becomes independent of $k_y$ at low $k_y$~\cite{Tang1980, Aleynikova2017, Pueschel2008}, as in Figure~6 of~\cite{kennedy2024} for STEP at reduced $\beta'$.}

At the operating point, equilibrium flow shear halves the growth rate at S60, but barely changes it at S50. Over the plane, flow shear removes the most weakly unstable regions but not the strongly driven ones. At S60 the static growth rate in the ideal-unstable region exceeds the shearing rate, and the Floquet averages keep about two thirds of the static value. At S50 the shearing rate exceeds every static growth rate, yet the Floquet-averaged growth rate along the ideal boundary remains between an eighth and a quarter of it. Flow shear widens the stable margin, but does not stabilise the ideal-unstable region. 

\subsubsection{Mode character}
\label{sec:salpha-mode-character}

\begin{figure}[tbp]
  \centering
  \includegraphics[width=\textwidth]{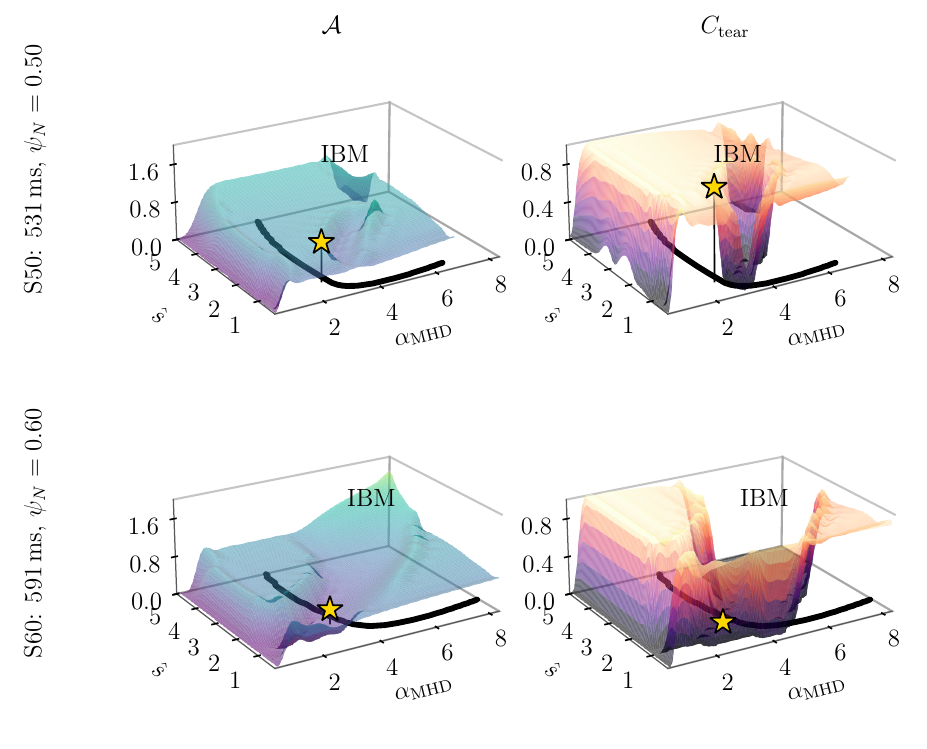}
  \caption{Electromagnetic amplitude ratio $\mathcal{A}$ of~(\ref{eq:ctear}), capped at $2$ (left-hand side), and tearing parameter $C_{\mathrm{tear}}$ of~(\ref{eq:ctear}) (right-hand side) of the $k_y\rho_s = 0.3$ mode in each cell of the static scans of Figure~\ref{fig:salpha_mountains}; rows S50, S60. A cell in which that mode is damped shows it all the same. Star, stem, axis ranges and $\gamma_{\mathrm{IBM}} = 0$ curve as in Figure~\ref{fig:salpha_mountains}.}
  \label{fig:salpha_em_maps}
\end{figure}

To identify the fastest-growing unstable mode (dominant) in each cell, we use the electromagnetic amplitude ratio $\mathcal{A}$ and the tearing parameter $C_{\mathrm{tear}}$~\cite{Patel2022},
\begin{equation}
  \mathcal{A} \equiv \left[\frac{\displaystyle\int|A_\parallel|^2\,\mathrm{d}\theta}{\displaystyle\int|\varphi|^2\,\mathrm{d}\theta}\right]^{1/2},
  \qquad
  C_{\mathrm{tear}} \equiv \frac{\left|\displaystyle\int A_\parallel\,\mathrm{d}l\right|}{\displaystyle\int|A_\parallel|\,\mathrm{d}l},
  \label{eq:ctear}
\end{equation}
where $l$ is the length along the field line. $C_{\mathrm{tear}}$ is near unity for tearing parity and near zero for ballooning parity. Figure~\ref{fig:salpha_em_maps} shows both over the static scans of Figure~\ref{fig:salpha_mountains}. On both surfaces the ideal-unstable region is strongly electromagnetic, with ballooning parity; at the operating magnetic shear, the mode returns to tearing parity still inside that region, above roughly $1.6$ times the operating $\alpha_{\mathrm{MHD}}$ at S50 and twice it at S60, where the growth rate has fallen again. Near its operating point S50 has tearing parity, whilst S60 has ballooning parity, just above the boundary between the two.

\subsection{Summary of linear results}
\label{sec:linear-summary}

In summary, both surfaces are IBM-stable, though sit close to the IBM stability boundary in $\alpha_{\mathrm{MHD}}$, with no second-stable region; the growth rate at $k_y\rho_s = 0.3$ rises towards the ideal boundary on both, with S60 on that rise and S50 on the microtearing plateau below it. The fastest-growing unstable mode (dominant) is a microtearing mode at S50 and a ballooning mode at S60. Equilibrium flow shear reduces the peak growth rate by about a fifth at S50 and three fifths at S60, but leaves the ideal-unstable region unstable. Importantly, both KBM-like modes and MTMs are unstable in both surfaces and they do not sit near an obvious threshold for the onset of any new linear instability: the growth rate is continuous across the ideal boundary (Figure~\ref{fig:salpha_mountains}).

In Section~\ref{sec:nonlinear}, we examine whether the turbulence at these surfaces saturates.

\section{Nonlinear simulations}
\label{sec:nonlinear}

In this section, we use nonlinear simulations to study turbulence on S50 and S60. Unstable ion-scale EM instabilities were identified in Section~\ref{sec:equilibrium-linear} and so it is interesting to ask where these surfaces lie relative to the non-zonal transition. As discussed in Section~\ref{sec:intro-48657}, each surface may lie below it, inside the no-go zone (facilitated by flow shear), or in the second-stable regime. 

\subsection{Numerical setup and the two nonlinear states}
\label{sec:nl-setup}

Our nonlinear simulations reported in this section retain two kinetic species, all three fields, and the full-plasma $\beta'$, but no equilibrium flow except where explicitly reported. The simulations resolve only the ion scale modes, $k_y\rho_s \le 0.92$ and use the numerical grids discussed in Table~\ref{tab:numerics}. We begin by exploring how the turbulence changes as the plasma becomes increasingly electromagnetic. At each surface, $\beta_e$, $\beta'$ and $\alpha_{\mathrm{MHD}}$ are scaled together by a multiplier $m_\beta$ from half to one and a half times their experimental values.

\texttt{GENE} gives the heat flux $Q$, summed over the kinetic species and over the electrostatic and electromagnetic channels, in gyro-Bohm units, $Q_{\mathrm{gB}} = n_e T_e c_s \rho_*^2$ with $\rho_* = \rho_s/a$, as the flux-surface average of $\bm{Q}\cdot\nabla x$, where $\bm{Q}$ is the heat-flux density and $x$ the radial coordinate. We divide it by $\langle|\nabla x|\rangle$ to give a flux per unit area, the same object calculated by \texttt{TRANSP}, and compare it with $Q_{\mathrm{TRANSP}} = 0.082$ and $0.094\,\mathrm{MW\,m^{-2}}$ at S50 and S60 (Section~\ref{sec:transp-reconciliation}). The simulations of this and the next three subsections have no equilibrium flow shear, which the experiment shows to be important; it is added in Section~\ref{sec:nl-flowshear}. We measure zonal regulation by the ratio
\begin{equation}
  \frac{|\varphi_{\mathrm{NZ}}|^2}{|\varphi_{\mathrm{ZF}}|^2} \equiv
  \frac{\displaystyle 2\sum_{k_y > 0}\sum_{k_x}\overline{\left\langle|\varphi_{\mathbf{k}}|^2\right\rangle}}
       {\displaystyle \sum_{k_x}\overline{\left\langle|\varphi_{k_x,\,k_y = 0}|^2\right\rangle}},
  \label{eq:nz-zf}
\end{equation}
where $\varphi_{\mathbf{k}}$ is the Fourier amplitude of the electrostatic potential at $\mathbf{k} = (k_x,\,k_y)$, $\langle\cdot\rangle$ the flux-surface average and the overbar the window average; NZ and ZF denote the non-zonal ($k_y \neq 0$) and zonal ($k_y = 0$) parts.

We classify each simulation by two conditions, both tested over its last tenth: (i) the mean heat flux exceeds $4\,Q_{\mathrm{TRANSP}}$; and (ii) the ratio in~(\ref{eq:nz-zf}), evaluated at each time rather than averaged, exceeds unity at least once, i.e., the non-zonal $|\varphi|^2$ reaches the zonal. A simulation that meets \emph{both} conditions is in the \emph{high-flux state}; otherwise it is in the \emph{low-flux state}. Condition (ii) tests for the signature of the non-zonal transition, the non-zonal potential overtaking the zonal one (Section~\ref{sec:nl-betascan}). It separates a large flux with weak zonal flows from a large flux that the zonal flows still regulate, such as that of S60 at its experimental $\beta_e$, which is several times power balance with a dominant zonal potential.\footnote{We note that the factor of four in condition (i) is not critical: no state changes for any factor between $3.5$ and $20$. The four-species S50 simulation at the experimental $\beta_e$ lies closest to the threshold (Section~\ref{sec:species-nonlinear}).} A simulation in the low-flux state runs to $t = 500\,a/c_s$. A simulation whose flux grows without turning over is stopped once its state is unambiguous. The fluxes and ratios shown in the figures are averaged over the last fifth of each simulation.

\subsection{Nonlinear gyrokinetic $\beta_e$ scan}
\label{sec:nl-betascan}

\begin{figure}
  \centering
  \includegraphics[width=\textwidth]{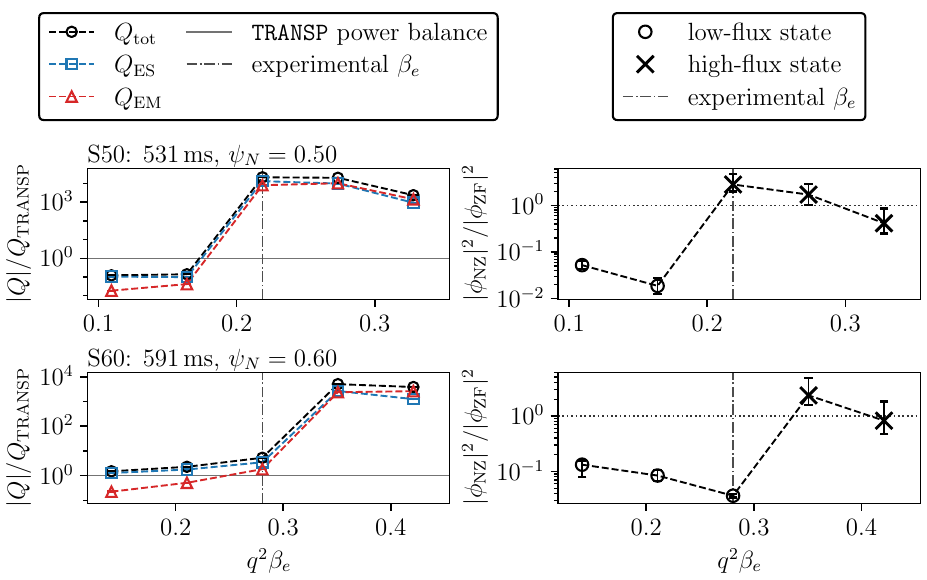}
  \caption{Heat flux $|Q|$ summed over species (left-hand side), total (black circles), electrostatic (\textcolor{tabblue}{blue}  squares) and electromagnetic (\textcolor{tabred}{red} triangles), and $|\varphi_{\mathrm{NZ}}|^2/|\varphi_{\mathrm{ZF}}|^2$ of~(\ref{eq:nz-zf}) with 16th to 84th percentile bars (right-hand side) against $q^2\beta_e$. All markers are open. On the right-hand side, circles denote the low-flux state and crosses the high-flux state; dash-dot vertical, the experimental $\beta_e$. The heat flux is shown in units of $Q_{\mathrm{TRANSP}}$, the total \texttt{TRANSP} power-balance heat flux, electron plus ion, with $Q_{\mathrm{TRANSP}} = 0.082$ and $0.094\,\mathrm{MW\,m^{-2}}$ at S50 and S60. The grey line is $Q_{\mathrm{TRANSP}}$.}
  \label{fig:nonlinear-betascan}
\end{figure}

Figure~\ref{fig:nonlinear-betascan} shows the heat flux (left-hand side) and the ratio in~(\ref{eq:nz-zf}) (right-hand side) against $q^2\beta_e$. At S60 and its experimental $\beta_e$, the simulation (neglecting $\gamma_E$) saturates at roughly five times power balance. At half this $\beta_e$, the flux is about one and a half times power balance and still rising slowly, and from $m_\beta = 1.25$ it is in the high-flux state, nearly three orders of magnitude above the low-flux level. In the absence of equilibrium flow shear, S60 sits on the cusp of the non-zonal transition. Contrastingly, at S50, the flux is already in the high-flux state at the experimental $\beta_e$. The time-averaged non-zonal to zonal ratio is a few hundredths to about a tenth in the low-flux state, and of order unity in the high-flux state.

\begin{figure}
  \centering
  \includegraphics[width=\textwidth]{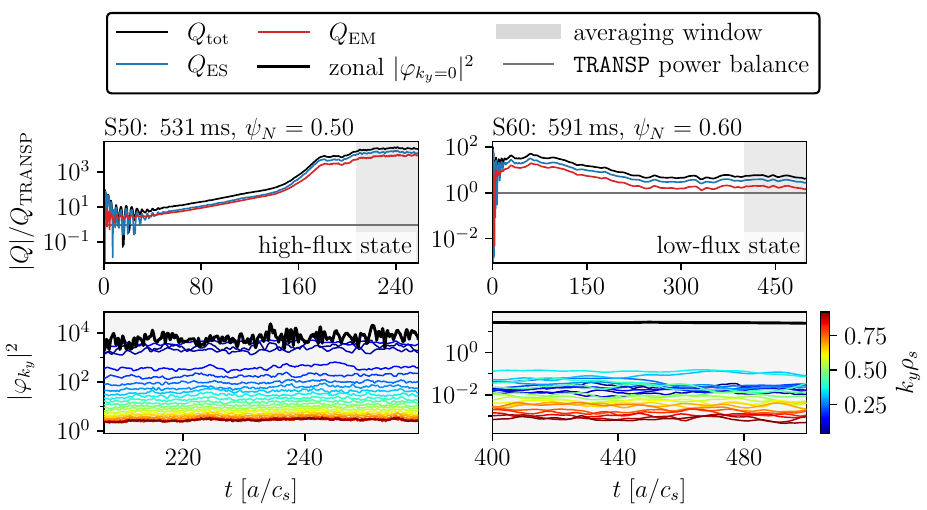}
  \caption{Heat flux against time over the whole simulation (top row) and $|\varphi_{k_y}|^2$ of every binormal mode against time over the averaging window (bottom row) at the experimental $\beta_e$ of each surface (columns), with its state (Section~\ref{sec:nl-setup}) printed in each flux panel. Top row: heat flux in units of $Q_{\mathrm{TRANSP}}$, as in Figure~\ref{fig:nonlinear-betascan}; the total flux is black, the electrostatic flux \textcolor{tabblue}{blue} and the electromagnetic flux \textcolor{tabred}{red}; the shading marks the averaging window, and the grey line is $Q_{\mathrm{TRANSP}}$. Bottom row: $k_y\rho_s$ is given by the colour bar, and the zonal mode is shown in black.}
  \label{fig:nl-stress-traces}
\end{figure}

Figure~\ref{fig:nl-stress-traces} shows the heat flux and $|\varphi_{k_y}|^2$ against time at the experimental $\beta_e$. At S60, the zonal mode remains two orders of magnitude above every finite-$k_y$ mode, whereas at S50, in the high-flux state, the longest wavelength reaches twice the zonal mode. Drift waves outgrowing the zonal mode at low $k_y$ is the key signature of the non-zonal transition~\cite{Pueschel2008, kennedy2026}.

\subsection{Nonlinear gyrokinetic simulations in the $(|q|,\,\beta_e)$ plane}
\label{sec:nl-qbeta}

\begin{figure}
  \centering
  \includegraphics[width=\textwidth]{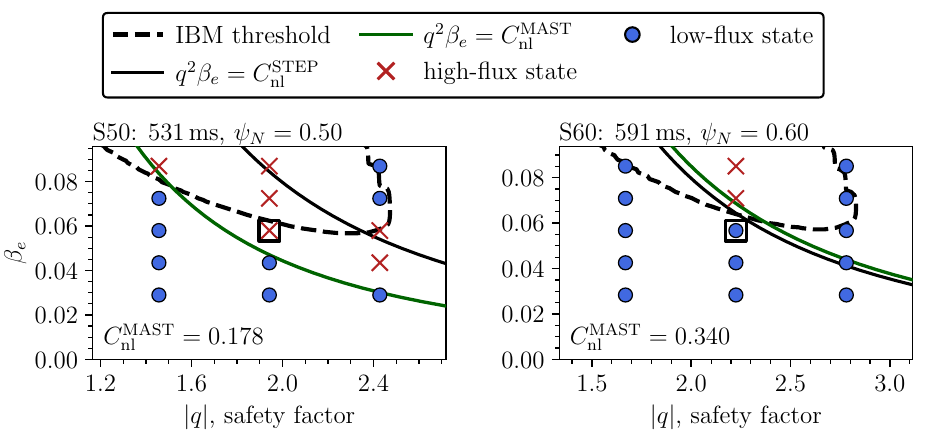}
  \caption{Outcome of every nonlinear simulation on the $(|q|,\,\beta_e)$ plane, one panel per surface, after Figure~3 of~\cite{kennedy2026}: the $\beta$ scan of Figure~\ref{fig:nonlinear-betascan} at multipliers $m_q \in \{0.75,\,1,\,1.25\}$ on $q$. Simulations in the high-flux state are marked by \textcolor{red}{red crosses} and those in the low-flux state by \textcolor{blue}{blue circles} (see text); the open black square is the experimental operating point. The solid black line is $q^2\beta_e = C_{\mathrm{nl}}^{\mathrm{STEP}}$ of~\cite{kennedy2026}, the \textcolor{ForestGreen}{green} line is $q^2\beta_e = C_{\mathrm{nl}}^{\mathrm{MAST}}$ (see text), and the dashed black line is the IBM boundary, with $q$ and $\beta'$ varied at the surface's own $\hat{s}$ and shape.}
  \label{fig:qbeta-grid}
\end{figure}

Figure~\ref{fig:qbeta-grid} shows the results of nonlinear gyrokinetic simulations on the $(|q|,\,\beta_e)$ plane of~\cite{kennedy2026}. Here, along with $\beta_e,$ the safety factor $q$ is varied by a multiplier $m_q$ at fixed $\beta'$ and $\hat{s}$, so that $\alpha_{\mathrm{MHD}}$ scales as $q^2$. The IBM boundary is a closed curve (dashed black) in this plane. As expected, every simulation well inside the IBM unstable region is in the high-flux state. On both surfaces, the gyrokinetic second-stable edge can be identified at $m_q = 1.25$ where the simulations saturate in the low-flux state at larger values of $\beta_e$. Following~\cite{kennedy2026}, we take $C_{\mathrm{nl}}^{\mathrm{MAST}}$ as the $q^2\beta_e$ contour that misclassifies the fewest simulations below 2nd stability on each surface. We find that $C_{\mathrm{nl}}^{\mathrm{MAST}} = 0.340$ at S60, close to the value $C_{\mathrm{nl}}^{\mathrm{STEP}} = 0.32$ at the STEP surface of~\cite{kennedy2026}, but $C_{\mathrm{nl}}^{\mathrm{MAST}}=0.178$ at S50. As expected from Figure~\ref{fig:nonlinear-betascan}, the two-species S60 operating point therefore lies below the transition, and the two species S50 operating point inside the no-go zone. Again, we stress that these calculations are without including the strong rotational shear in MAST-U; we do not expect fluxes from these simulations to agree with those from experiment, but take them instead to indicate proximity of MAST-U to the NZT if flow shear were absent. The S60 value of $C_{\mathrm{nl}}^{\mathrm{MAST}}$ suggests that MAST-U can approach the non-zonal transition at STEP-like values of $q^2\beta_e$.

\subsection{Linear eigenvalues over the $(|q|,\,\beta_e)$ plane}
\label{sec:nl-qbeta-linear}

\begin{figure}
  \centering
  \includegraphics[width=\textwidth]{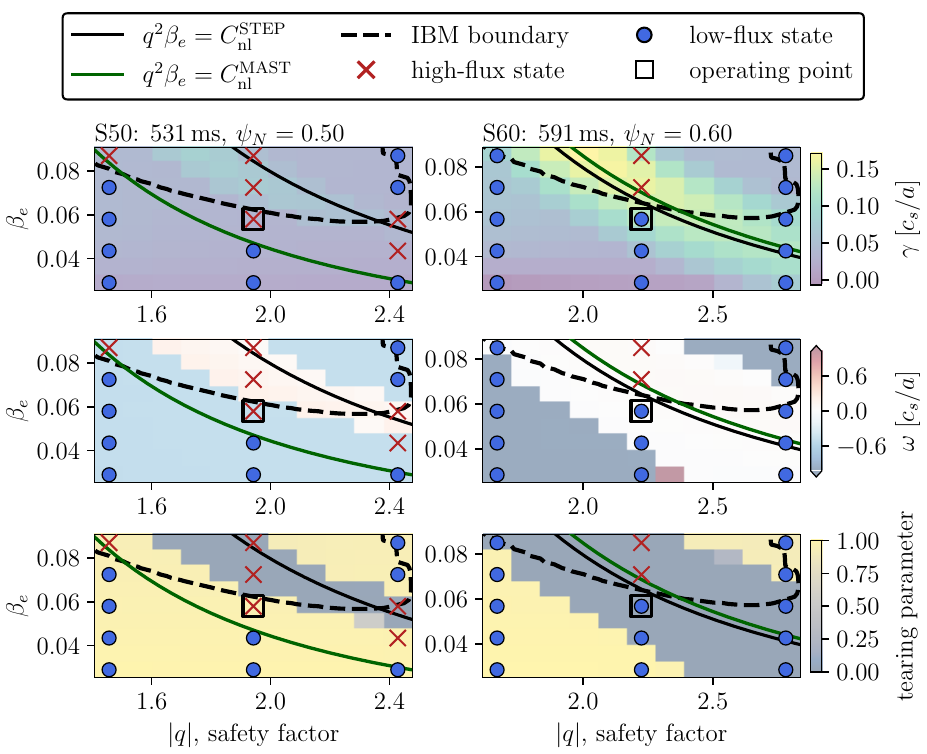}
  \caption{Growth rate, real frequency (positive in the ion direction) and $C_{\mathrm{tear}}$ of~(\ref{eq:ctear}) (rows) over the $(|q|,\,\beta_e)$ plane of Figure~\ref{fig:qbeta-grid} at each surface's growth-rate peak, $k_y\rho_s = 0.31$ and $0.46$ (columns). Markers and curves are as in Figure~\ref{fig:qbeta-grid}.}
  \label{fig:qbeta-linear-peak}
\end{figure}

Figure~\ref{fig:qbeta-linear-peak} shows the linear eigenvalues of the fastest-growing unstable mode (dominant) over the same plane at each surface's growth-rate peak. There is no obvious dramatic change in the linear physics across the transition threshold in agreement with \cite{kennedy2026}. At S60, $\beta^\prime$-stabilisation reduces the growth rate just above the nominal $q$ at $m_\beta = 1.25$, so the $m_q = 1.25$ simulations that saturate above the boundary are also more weakly driven. The S60 parity changes from tearing to ballooning just below the operating point, and back to tearing at the largest drive.

\subsection{Field-line stochasticisation}
\label{sec:nl-stochastic}

\begin{figure}
  \centering
  \includegraphics[width=\textwidth]{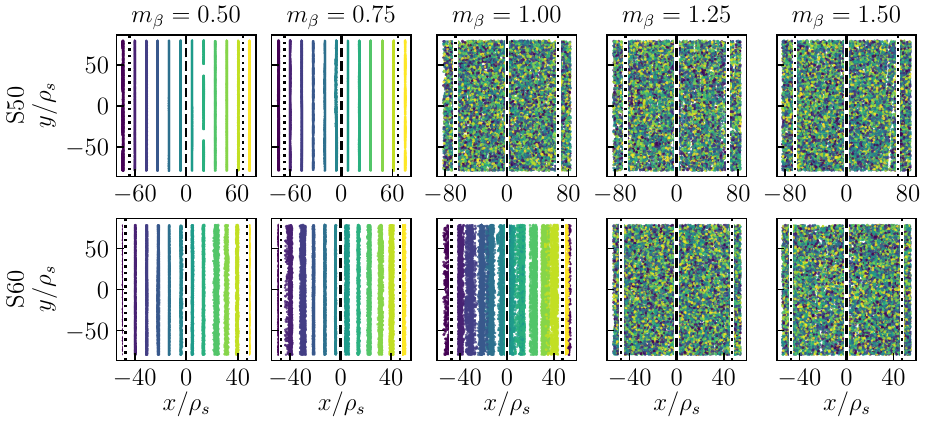}
  \caption{Poincar\'e maps of the perturbed field at the outboard midplane, coloured by launch radius. Rows are surfaces (S50 at $531\,\mathrm{ms}$ and $\psi_N = 0.50$, S60 at $591\,\mathrm{ms}$ and $\psi_N = 0.60$) and columns are $\beta$ multipliers. Black dashed and dotted verticals, edged in white, mark the rational surfaces of the two lowest finite-$k_y$ modes, of spacing $2\pi/(\hat{s}k_{y,\min})$ and half that, where $k_{y,\min}$ is the smallest finite $k_y$.}
  \label{fig:poincare}
\end{figure}

It is revealing to connect the non-zonal transition directly to stochasticisation of the magnetic field. Figure~\ref{fig:poincare} shows Poincar\'e maps across the $\beta_e$ scan. These maps are constructed in \texttt{pyrokinetics} by following magnetic field lines for many toroidal turns and recording their successive intersections with a fixed poloidal plane: intact flux surfaces therefore appear as smooth, nested curves, whereas stochastic field lines wander radially and progressively fill a finite volume. We classify a surface as closed when the 5th--95th percentile of its radial excursion remains smaller than the spacing between neighbouring launch surfaces.

The change at the experimental $\beta_e$ on the S50 surface (inside the no-go zone) is striking. Below this value, every launched field line remains confined to a closed surface, giving the ordered, laminar topology expected for nested magnetic surfaces. At and above the experimental $\beta_e$, those surfaces are destroyed and the intersections become volume-filling, indicating a transition to fully mixed magnetic chaos rather than merely the appearance of isolated stochastic layers associated with MTM physics (e.g.,~\cite{Giacomin2023microtearing}). This distinction is important: a stochastic layer consists of a radially limited chaotic region, typically bounded by surviving flux surfaces, so that field-line mixing remains local. Here, by contrast, the destruction of the nested surfaces allows chaotic field lines to explore essentially the full accessible radial volume, opening a much more effective channel for parallel electron transport across the plasma. Notably, for S60 at its experimental $\beta_e$, roughly two thirds of the launched surfaces remain closed even though the simulated flux slightly exceeds power balance (i.e., complete stochasticisation of the magnetic field lines is another way of classifying points above the non-zonal transition from those below it).

This topological transition provides a natural interpretation of the accompanying non-zonal transport. Reduced models have found stochastic-field electron transport to be important in MAST and MAST-U~\cite{Palermo2025}, while nonlinear microtearing simulations of both MAST and NSTX find that electron heat transport can be carried predominantly by magnetic flutter~\cite{Giacomin2023microtearing, patel2025exb, singh2026}. The Poincar\'e maps therefore suggest that the sharp transport transition is associated not simply with stronger magnetic perturbations, but with a qualitative change in field-line topology: from laminar, nested flux surfaces to volume-filling stochasticity.

\subsection{Equilibrium flow shear}
\label{sec:nl-flowshear}

\begin{figure}
  \centering
  \includegraphics[width=\textwidth]{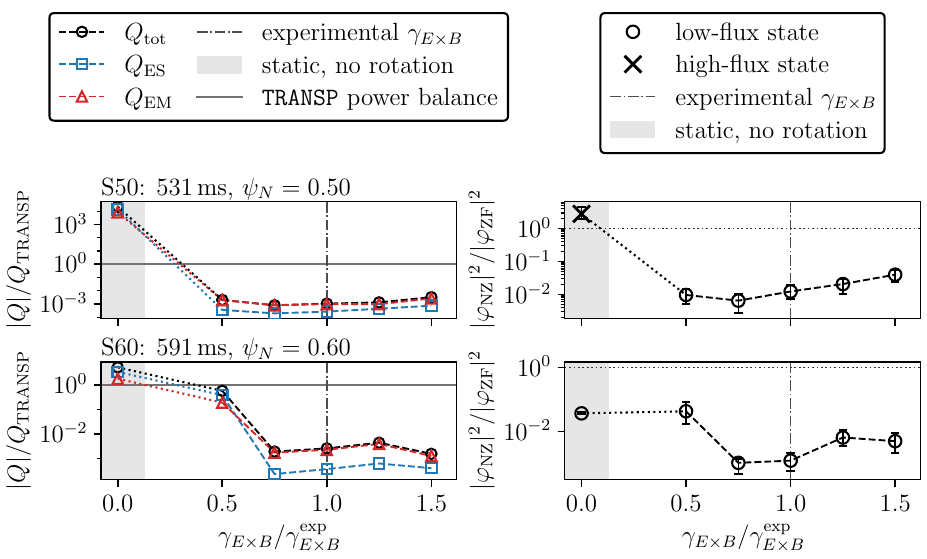}
  \caption{Heat flux $|Q|$ (left-hand side) and $|\varphi_{\mathrm{NZ}}|^2/|\varphi_{\mathrm{ZF}}|^2$ (right-hand side) against $\gamma_{E\times B}$ over its experimental value (dash-dot vertical) at fixed $\beta_e$, $\beta'$ and $\alpha_{\mathrm{MHD}}$, one row per surface. Colours, markers, the grey line and the units of heat flux are as in Figure~\ref{fig:nonlinear-betascan}. The grey band at $\gamma_{E\times B} = 0$ marks the static simulations of Figure~\ref{fig:nonlinear-betascan}, without equilibrium rotation, which are joined to the finite-shear points by dotted lines.}
  \label{fig:nl-flowshear}
\end{figure}

Figure~\ref{fig:nl-flowshear} repeats Figure~\ref{fig:nonlinear-betascan} as a scan in $\gamma_{E\times B}$, with the Coriolis and centrifugal terms retained so that the gyrokinetic equations are solved in a rotating reference frame. At S60, the flux drops from several times power balance in the absence of flow shear to roughly three fifths of power balance at half the experimental $\gamma_{E\times B}$, and to more than two orders of magnitude below power balance once the shearing rate reaches three quarters of the experimental value. At S50, every simulation with finite flow shear lies more than two orders of magnitude below power balance. Thus, with only two kinetic species, the experimental shearing rate suppresses the turbulent flux far below power balance at both surfaces. Most strikingly, S50---in the high-flux state without flow shear---is apparently held inside the no-go zone by the experimental shear.

\begin{figure}
  \centering
  \includegraphics[width=\textwidth]{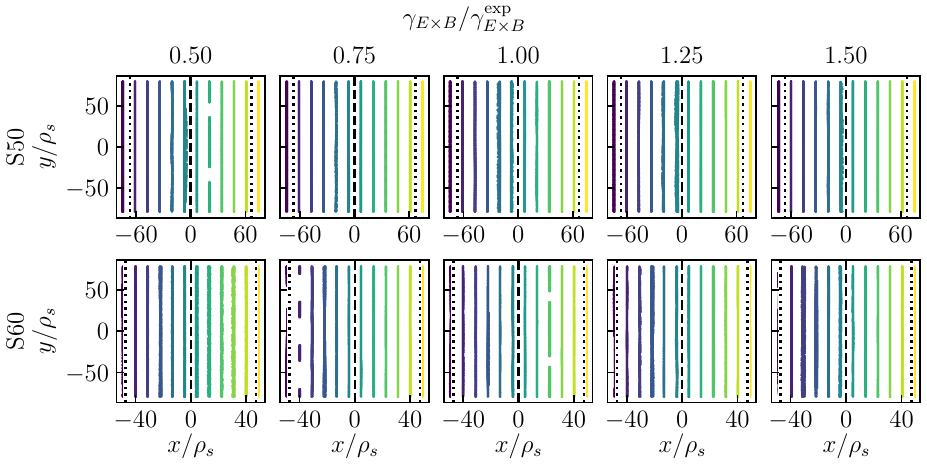}
  \caption{Poincar\'e maps as in Figure~\ref{fig:poincare} across the flow-shear scan of Figure~\ref{fig:nl-flowshear}; rows are surfaces (S50 at $531\,\mathrm{ms}$ and $\psi_N = 0.50$, S60 at $591\,\mathrm{ms}$ and $\psi_N = 0.60$) and columns are $\gamma_{E\times B}$ over its experimental value. Black dashed and dotted verticals, edged in white, mark the rational surfaces as in Figure~\ref{fig:poincare}. The simulations without flow shear are not repeated here: they are the $m_\beta = 1$ column of Figure~\ref{fig:poincare}.}
  \label{fig:poincare-flowshear}
\end{figure}

The magnetic topology changes at the same time (Figure~\ref{fig:poincare-flowshear}). In the sheared simulations the field lines are again laminar, remaining confined to intact flux surfaces rather than wandering stochastically through the radial domain. The suppression of transport is therefore accompanied by a qualitative re-ordering of the magnetic field: flow shear removes the volume-filling stochasticity seen in the unsheared state and restores nested magnetic surfaces. This suggests that the reduction in electron heat flux is tied not only to decorrelation of the turbulence, but also to the loss of the stochastic magnetic transport channel.

The apparent over-suppression is, however, at least partly a consequence of the reduced species model. When carbon is included, in the three- and four-species simulations, the S60 flux at the experimental shearing rate rises back to order power balance (Section~\ref{sec:species-nonlinear}). Why the two-species set underestimates the sheared transport is not resolved; Section~\ref{sec:species-nonlinear} lists the candidates: the lower electron density gradient of the two-species set, the impurity shielding of the potential, and the electron collisions with carbon (and the impact of collisionality on the zonal flow).

\section{The role of impurities and fast ions in the ion-scale transport}
\label{sec:species}

The plasma of \#48657 also contains carbon and fast beam deuterium, and both carbon~\cite{singh2026} and fast ions~\cite{Dudkovskaia2026} can change the simulated transport in STs. Since the non-zonal transition rests on a balance between the zonal flows and the turbulence, to which each species contributes with its own weight, the central message of Section~\ref{sec:nonlinear}---that NBI-driven flow shear decides whether the ion scale carries the anomalous flux---ought to be tested against the most complete multi-species local description of the plasma. 

\subsection{The species sets}
\label{sec:species-sets}

We compare the two-species set used so far with three species (electrons, thermal deuterium, and carbon) and four (with the fast beam deuterium), both linearly and nonlinearly.

\subsubsection{Which species can be dropped}
\label{sec:species-two}

The two-species reduction~(\ref{eq:species_two}) of Section~\ref{sec:linear-species} keeps $\beta$, $\beta'$ and $\alpha_{\mathrm{MHD}}$ at their four-species values of~(\ref{eq:app_two_sums}); the four-species \texttt{TRANSP} values from which every reduced set is built are in Table~\ref{tab:species_inputs}. Whether FLR effects can justify dropping a species from the kinetic description depends on its Larmor radius relative to the ion scale,
\begin{equation}
  \frac{\rho_f}{\rho_s} = \frac{1}{Z_f}\sqrt{\frac{m_f T_f}{m_i T_e}}
  = \sqrt{\frac{T_f}{T_e}},
  \qquad
  \frac{\rho_C}{\rho_s} = \frac{1}{Z_C}\sqrt{\frac{m_C T_C}{m_i T_e}},
  \label{eq:app_larmor}
\end{equation}
where $m_f = m_i$ and $Z_f = 1$ for the beam deuterium, and $m_C$ and $T_C = T_i$ are the mass and temperature of carbon. At S50, $T_f \approx 33\,T_e$ gives $\rho_f \approx 6\,\rho_s$, so that $k_\perp\rho_f$ is several times unity at the spectral peak, where the gyroaverage
\begin{equation}
  \Gamma_0(b) = I_0(b)\,e^{-b}, \qquad b = k_\perp^2\rho_f^2,
  \label{eq:app_gyroaverage}
\end{equation}
reduces their response to a few tens of per cent at most. Here, $I_0$ is the modified Bessel function and $k_\perp$ the wavenumber perpendicular to the field. The fast-ion orbits also span a sizeable fraction of the minor radius. The local ordering requires the Larmor radius $\rho_{\mathrm{L},s} = \sqrt{m_s T_s}/(|Z_s| e B_0)$ of each species to be small compared with the minor radius, $\rho_{*,s} = \rho_{\mathrm{L},s}/a \ll 1$. Table~\ref{tab:species-rhostar} gives $\rho_{*,s}$ for every species, from the four-species inputs of Table~\ref{tab:species_inputs}: it is below a fiftieth for electrons, deuterium and carbon, but about six times larger for the fast ions, $\rho_{*,f} \approx 0.08$. A sharper measure compares the radial excursion of a trapped orbit, the banana width $\Delta_{b,s} \approx |q|\rho_{\mathrm{L},s}/\sqrt{\epsilon}$ with $\epsilon = r/R$ the inverse aspect ratio, with the length $L_{n,s}$ of the species' own density gradient. This ratio is a few hundredths for the bulk species, but exceeds unity for the strongly peaked fast ions (Table~\ref{tab:species-rhostar}): their orbits are wider than their density gradient scale length, so challenge the gyrokinetic ordering. That the ordering is broken for the fast ions does not mean that they can be ignored: their pressure stays in the equilibrium through $\beta'$, and their kinetic response is tested below with four species, with the caveat that a local model may describe it poorly. Carbon has $\rho_C \approx 0.5\,\rho_s$: it sits at the ion scale and responds fully there, so its kinetic response is expected. We proceed to test both reductions explicitly with three and four species.

\begin{table}[tbp]
  \caption{Normalised Larmor radius $\rho_{*,s} = \rho_{\mathrm{L},s}/a$, with $\rho_{\mathrm{L},s} = \sqrt{m_s T_s}/(|Z_s| e B_0)$, and banana width $\Delta_{b,s} = |q|\rho_{\mathrm{L},s}/\sqrt{\epsilon}$ over the density gradient length $L_{n,s}$ of each species at the two analysis surfaces, from the four-species \texttt{GENE} inputs of Table~\ref{tab:species_inputs}.}
  \label{tab:species-rhostar}
  \centering
  \begin{minipage}{0.85\textwidth}
  \begin{tabular*}{\linewidth}{@{\extracolsep{\fill}}lcccc@{}}
\toprule
 & \multicolumn{2}{c}{$\rho_{*,s}$} & \multicolumn{2}{c}{$\Delta_{b,s}/L_{n,s}$} \\
\cmidrule(lr){2-3} \cmidrule(l){4-5}
species & S50 & S60 & S50 & S60 \\
\midrule
electrons ($e$) & 0.0002 & 0.0002 & 0.00 & 0.00 \\
thermal deuterium ($i$) & 0.0166 & 0.0155 & 0.04 & 0.06 \\
carbon ($C$) & 0.0068 & 0.0063 & 0.02 & 0.03 \\
fast beam deuterium ($f$) & 0.0828 & 0.0857 & 1.22 & 1.89 \\
\bottomrule
\end{tabular*}

  \end{minipage}
\end{table}

\subsubsection{The three- and four-species sets}
\label{sec:species-three-four}

The three-species set removes only the fast deuterium,
\begin{equation}
  n_i^{(3)} = n_e - Z_C n_C,
  \qquad
  \left(\frac{a}{L_{n_i}}\right)^{(3)}
  = \frac{n_e\,a/L_{n_e} - Z_C n_C\,a/L_{n_C}}{n_e - Z_C n_C},
  \label{eq:species_three}
\end{equation}
which is~(\ref{eq:species_qn}) without the fast ions; the electron density and temperature, carbon, $\beta_e$, $\beta'_{\mathrm{full}}$ and $\alpha_{\mathrm{MHD}}$ are those of the four-species set, so that the two reduced sets differ in $a/L_{n_e}$ as well as in carbon. Unlike the two-species set, the three-species set keeps the \texttt{TRANSP} electron density gradient; the thermal deuterium gradient absorbs the loss of the fast ions. The four-species set is the \texttt{TRANSP} set, with the fast deuterium a Maxwellian at $T_f$. Figure~\ref{fig:species-cartoon} compares the three sets. In all three sets, $\beta'$ is the full-plasma \texttt{TRANSP} value $\beta'_{\mathrm{full}}$ of~(\ref{eq:app_two_sums}), including the fast-ion and rotation pressure, so that the sets differ only in the species evolved.

\subsection{Linear spectra}
\label{sec:species-linear}

\begin{figure}[tbp]
  \centering
  \includegraphics{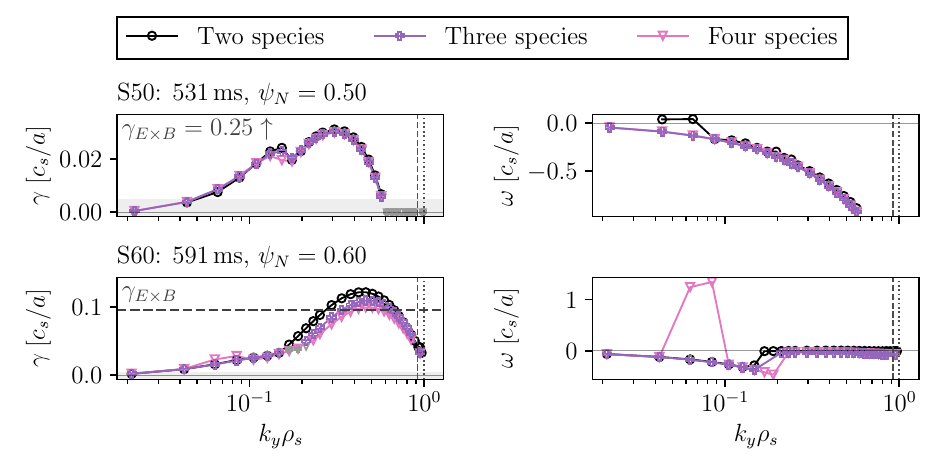}
  \caption{Growth rate (left-hand side) and real frequency, positive in the ion direction (right-hand side), of the fastest-growing unstable mode (dominant) against $k_y\rho_s$ at the full-plasma $\beta'$ without equilibrium flow, one row per surface, with two (black circles), three (\textcolor{tabpurple}{purple plus signs}), and four (\textcolor{tabpink}{pink triangles}) kinetic species. All markers are open: full-size markers in the curve colour, converged growing modes; small grey markers, damped or unconverged modes; the curves join the full-size markers only. Grey band, dashed horizontal and verticals as in Figure~\ref{fig:spectra}.}
  \label{fig:species-234-linear}
\end{figure}

Figure~\ref{fig:species-234-linear} shows the growth rate and real frequency of the fastest-growing unstable mode (dominant) with two, three, and four species. The shape of the spectra and the peak wavenumber are the same in all three cases. The peaks are somewhat lower with more species, as the dilution of the bulk ions implies, and at S60 the three- and four-species modes propagate in the electron direction whereas the two-species mode barely propagates; the two sets differ in the electron density gradient, which the two-species reduction lowers (Table~\ref{tab:species_inputs}), as well as in carbon, and we have not separated the two. Linearly, the additional species are a modest dilution of the drive, and nothing in these spectra anticipates a qualitative change in the nonlinear transport. This differs from the NSTX microtearing modes of~\cite{singh2026}, which carbon destabilises, probably through the electron collisions with carbon; at S50, carbon instead lowers the microtearing peak slightly, by a few per cent.

\subsection{Nonlinear fluxes}
\label{sec:species-nonlinear}

\begin{figure}[tbp]
  \centering
  \includegraphics[width=\textwidth]{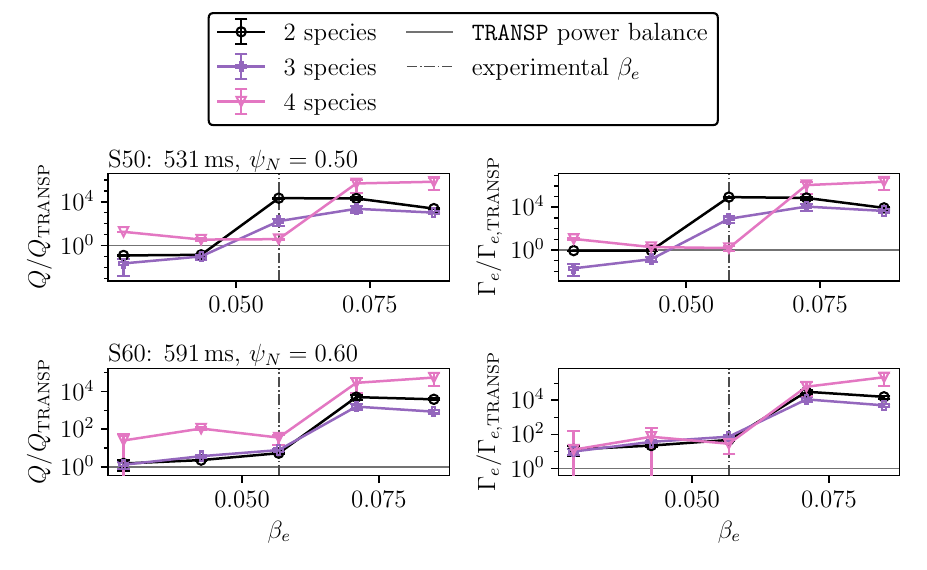}
  \caption{Heat flux summed over the kinetic species (left-hand side) and electron particle flux (right-hand side), electrostatic plus electromagnetic, against $\beta_e$, one row per surface, with two (black open circles), three (\textcolor{tabpurple}{purple plus signs}), and four (\textcolor{tabpink}{pink open triangles}) kinetic species. Bars, 16th to 84th percentile over the averaging window. Fluxes in units of their \texttt{TRANSP} references, $Q_{\mathrm{TRANSP}}$ as in Figure~\ref{fig:nonlinear-betascan} and the electron particle flux its sources imply in steady state, $\Gamma_{e,\mathrm{TRANSP}} = 1.2$ and $1.5\times10^{19}\,\mathrm{m^{-2}\,s^{-1}}$; grey line, the reference. Axes logarithmic, or symmetric-log where the quantity changes sign. Where a simulation ends before $t = 500\,a/c_s$, its final time is given.}
  \label{fig:flux-species}
\end{figure}

Figure~\ref{fig:flux-species} repeats the nonlinear $\beta$ scan neglecting toroidal rotation, of Section~\ref{sec:nonlinear}, with three and four species. At S50 and the experimental $\beta_e$, the heat flux is in the high-flux state with two and three species, whereas with four species it lies just below the threshold of our criterion and enters the high-flux state at the next value of $\beta_e$: the delay comes with the introduction of fast deuterium and not with the carbon, consistent with~\cite{Dudkovskaia2026}. At S60, the three-species fluxes follow the two-species ones and enter the high-flux state at the same $\beta_e$, whereas with four species the flux is already in the high-flux state at the experimental $\beta_e$. The species set thus moves each surface relative to the transition by about as much as its distance from it, in opposite directions at the two surfaces, but without flow shear every species set leaves at least one surface in the high-flux state at its experimental $\beta_e$. The question of how \#48657 avoids the no-go zone cannot be answered by the species physics; it must be answered by the rotation.

\begin{figure}[tbp]
  \centering
  \includegraphics[width=\textwidth]{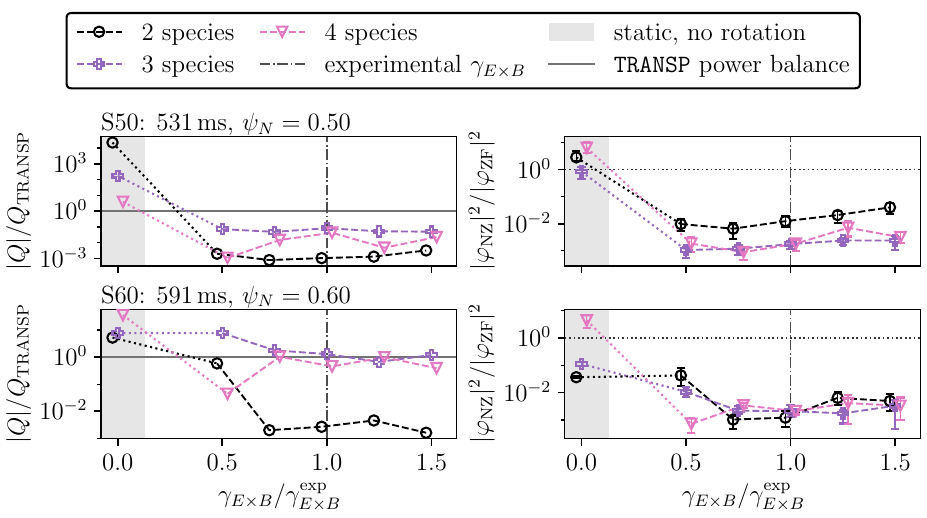}
  \caption{As Figure~\ref{fig:nl-flowshear}, with three (electrons, thermal deuterium, and carbon) and four kinetic species, one row per surface: total heat flux $|Q|$ in units of $Q_{\mathrm{TRANSP}}$ of Figure~\ref{fig:nonlinear-betascan} (left-hand side) and $|\varphi_{\mathrm{NZ}}|^2/|\varphi_{\mathrm{ZF}}|^2$ (right-hand side) with two (black open circles), three (\textcolor{tabpurple}{purple plus signs}), and four (\textcolor{tabpink}{pink open triangles}) kinetic species. The two-species points are those of Figure~\ref{fig:nl-flowshear}. Grey band, the static $1.00\,\beta_e$ simulations of Figure~\ref{fig:flux-species} at $\gamma_{E\times B} = 0$, joined to the finite-shear points by dotted lines. At S50 the three- and four-species fluxes with flow shear are negative (inward) but the plot shows only their magnitude.}
  \label{fig:flowshear-34species-paper}
\end{figure}

Figure~\ref{fig:flowshear-34species-paper} therefore repeats the flow-shear scan of Figure~\ref{fig:nl-flowshear} with three and four species. At S50, flow shear suppresses the flux with every species set to well over an order of magnitude below power balance. Only a very modest fraction of the anomalous flux at S50 can be attributed to ion-scale turbulence in the presence of the measured rotation. At S50 the three- and four-species points lie above the two-species ones although their $|\varphi_{\mathrm{NZ}}|^2/|\varphi_{\mathrm{ZF}}|^2$ is lower. This does not mean stronger turbulence. The panels show the magnitude $|Q|$ of the net heat flux, and with three and four species that net flux is negative, a few hundredths to a tenth of power balance: an inward electromagnetic (magnetic-flutter) flux outweighs a smaller outward electrostatic one. The two-species flux is outward but smaller still. Comparing $|Q|$ between the species sets at S50 therefore compares the size of small residual fluxes of opposite sign, not the strength of the turbulence. At S60, the three- and four-species fluxes with flow shear lie within a factor of a few of power balance, whereas with two species they lie orders of magnitude below it. With the experimental flow shear, the S60 heat flux therefore reaches power balance only when carbon is included. The apparent over-suppression of the two-species simulations was a consequence of the reduced species model: the rotation does not remove the ion-scale turbulence at S60, but holds it at the level the power balance requires, whilst the same surface without rotation carries a flux more than an order of magnitude larger. The measured rotation keeps S60 on the low-flux side of the transition, with little margin. Why the two-species reduction underestimates the sheared transport at S60 is not resolved here. The candidates are the lower electron density gradient of the two-species set, the impurity shielding of the potential, the electron collisions with carbon, and a stronger collisional damping of the zonal flows through the higher collisionality that carbon brings, as found for ETG turbulence~\cite{Colyer2017}; separating them needs simulations that change one at a time.

A plausible reason for the stronger dependence of the nonlinear results on the species set is that the transition is set by the zonal flows~\cite{Zhang2025b, kennedy2026}. The electrostatic potential for any $k$ follows from gyrokinetic quasineutrality,
\begin{equation}
  \sum_s \frac{Z_s^2 e^2 n_s}{T_s}\,\varphi_{\mathbf{k}}
  = \sum_s Z_s e \int J_0\!\left(\frac{k_\perp v_\perp}{\Omega_s}\right) h_{s,\mathbf{k}}\,\mathrm{d}^3 v,
  \label{eq:species_qn_gk}
\end{equation}
where $h_s$ is the non-adiabatic part of the perturbed distribution, $J_0$ the Bessel function, $v_\perp$ the velocity perpendicular to the field and $\Omega_s$ the cyclotron frequency of species $s$. The Boltzmann shielding on the left-hand side weights each species by $Z_s^2 n_s/T_s$. Through $Z_C^2 = 36$, carbon, at under two per cent of the electron density, still contributes about a fifth of the shielding at S50, and so reduces the potential at every $k_\perp$, the zonal potential included. We speculate that carbon could move the transition in two ways: its shielding may act more on the zonal potential than on the non-zonal one, and its collisions with deuterium, weighted by $Z_C^2$, may damp the zonal flows more strongly, as found for ETG turbulence~\cite{Colyer2017}. We have not isolated either, and the simulations only partly support them, since at S50 the delay of the high-flux state is due to the fast deuterium and not to the carbon. For NSTX microtearing modes, Singh \textit{et al.}~\cite{singh2026} attribute the destabilising effect of carbon to impurity shielding of the electrostatic potential, together with the higher electron--ion collisionality that carbon brings through $Z_{\mathrm{eff}}$. In STEP, kinetic impurities and the fast-particle pressure reduce the transport with impurity enhancement of electron collisions playing a key role~\cite{Bokshi2025}, the opposite of what Figure~\ref{fig:flowshear-34species-paper} shows at S60, where adding carbon raises the heat flux at the experimental flow shear.

\FloatBarrier

\section{Conclusions}
\label{sec:conclusions}

This work has addressed two questions using local electromagnetic gyrokinetic simulations of discharge \#48657 at two IBM-stable surfaces: S50, close to peak $\beta_N$, and S60, at the maximum IBM-stable $q^2\beta_e$. Firstly, is ion-scale turbulence important in MAST-U, and needed to explain the anomalous transport where the electron scale falls short? Secondly, can MAST-U access the turbulent regime expected in the core of STEP? We answer the first in Section~\ref{sec:conclusions-shortfall} and the second in the sections that follow. 

\subsection{Ion-scale turbulence and the electron-scale shortfall}
\label{sec:conclusions-shortfall}

The companion electron-scale study~\cite{patel2026etg} finds that, at peak $\beta_N$ in \#48657, ETG modes are unstable only at $\psi_N = 0.5$, where they drive a heat flux several orders of magnitude below the experimental level, and are stable further out, including at $\psi_N = 0.6$. The anomalous electron heat flux of the high-performance phase is therefore not from electron-scales, and the first question of this work is whether ion-scale turbulence makes up the shortfall. At S60 it does: with the measured flow shear and the carbon impurity included, Figure~\ref{fig:flowshear-34species-paper} shows that the ion-scale heat flux is of the order of power balance. Ion-scale electromagnetic turbulence is thus important in MAST-U, and needed to explain the anomalous transport of this surface. At S50 it seemingly does not: the sheared ion-scale flux lies far below power balance with every species set, as does the electron-scale flux, so that the anomalous flux there lies outside both local descriptions; coherent $n = 2$ activity is found near S50 shortly after its analysis time (Section~\ref{sec:fom-mhd}). Our gyrokinetic simulations on both surfaces suggest that the level of the ion-scale flux is limited by the toroidal rotation: without rotation, the ion scale turbulence would carry far more than the available heating.

\subsection{MAST-U without rotation recovers the gyrokinetic physics of STEP}
\label{sec:conclusions-step}

Linear simulations find that the fastest-growing unstable mode (dominant) is a microtearing mode at S50 and an electromagnetic ballooning mode at S60, and the ballooning mode takes over at S50 as $q^2\beta_e$ rises (see Figure~\ref{fig:param_scans}). Ballooning-like modes and MTMs are unstable on both surfaces analysed and these are the instabilities of the STEP core~\cite{kennedy2023a, kennedy2026}, even though on MAST-U $\beta_e$ is only a fraction of its value in STEP.

Without equilibrium flow shear, Figure~\ref{fig:nonlinear-betascan} shows that the two surfaces lie on either side of the non-zonal transition and close to it.  Fluxes on these MAST-U surfaces enter the high-flux state above a threshold $q^2 \beta_e >C_{\mathrm{nl}}$, where the threshold in MAST, $C_{\mathrm{nl}}^{\mathrm{MAST}}$, is close to that reported for STEP surface of~\cite{kennedy2026}, whilst that of S50 is about half as large.  The S60 flux saturates above power balance at the experimental $q^2 \beta_e$ and is in the high-flux state at higher values, whilst S50 lies inside the no-go zone and is in the high-flux state at the experimental $q^2 \beta_e$. Above the transition the longest wavelengths outgrow the zonal potential and the magnetic field is volume-filling and stochastic. MAST-U therefore reproduces, at a fraction of the STEP $\beta_e$, the phenomenology on which the predictions for STEP rest~\cite{Giacomin2024, kennedy2026, kennedy2026global}.

In an experiment, the fluxes above the transition could not be sustained by the available heating. The profiles would instead relax until $q^2\beta_e$ fell back to the threshold, so that the non-zonal transition would act as a very stiff limit, pinning the profiles at $q^2\beta_e \approx C_{\mathrm{nl}}$ below the ideal $\beta$ limit. Were it not for the rotation, the \#48657 profiles would have to adjust for local fluxes to achieve power balance, and a more significant reduction in flux would be required at S50. 

\subsection{Rotation in MAST-U allows operation above the transition}
\label{sec:conclusions-rotation}

Equilibrium flow shear driven by NBI allows plasmas to operate above the transition threshold in $q^2 \beta_e$, which is so far only mapped for $\gamma_E=0$. Including experimental flow shear in local equilibrium models with various numbers of species, the S50 flux is far below power balance, so S50 is held inside the no-go zone by the rotation. At S60, the two-species flux falls from several times power balance to far below it at the experimental $\gamma_{E\times B}$, but including carbon gets much closer to power balance: the rotation holds S60 at the level the power balance requires, whereas without it the flux would be more than an order of magnitude larger. NSTX microtearing simulations have shown a similar pattern, recovering the measured electron heat flux only below the full shearing rate~\cite{Guttenfelder2011, Guttenfelder2012, patel2025exb} or with carbon included~\cite{singh2026}.

Linearly, flow shear only reduces the growth rates; nonlinearly, it moves the transition from $C_{\mathrm{nl}}$ to an effective $C_{\mathrm{nl}}'$ beyond the operating point (as illustrated in Figure~\ref{fig:nzt-cartoon}, top); this is nonlinear gyrokinetic analogue of the suggestion in~\cite{harrison2026} that NBI-driven flow shear gives MAST-U access to IBM-marginal $\beta_e$. The high performance of \#48657 is thus not evidence that the non-zonal transition is absent in MAST-U, but suggests that the momentum injected by the beams holds the plasma on its low-flux side. STEP, with only weak flow shear~\cite{tholerus2024}, might not have this margin. Singh \textit{et al.}~\cite{singh2026} reach a similar conclusion for low-collisionality NSTX plasmas: there, hybrid ballooning-parity modes drive very large fluxes at modest equilibrium flow shearing rate, but the experimental flow shear is large enough to suppress these modes, leaving the ion heat flux near neoclassical.

\subsection{Large-scale MHD, loss of rotation and the non-zonal transition: cause and effect}
\label{sec:conclusions-arrows}

The most obvious difficulty in mapping these results onto STEP is that MAST-U \#48657 is not quiescent to MHD. A $2/1$ neoclassical tearing mode forms, collapses the core rotation and ends the high-performance phase~\cite{harrison2026}, and coherent low-$|n|$ modes are present over much of the discharge (Section~\ref{sec:candidate-selection}). The experiment thus presents a coincidence of large-scale MHD, the loss of rotation and the loss of performance; our gyrokinetic simulations suggest that large electromagnetic transport can be caused by the loss of rotation, but do not explain the latter. 

\begin{figure}[tbp]
  \centering
  \includegraphics{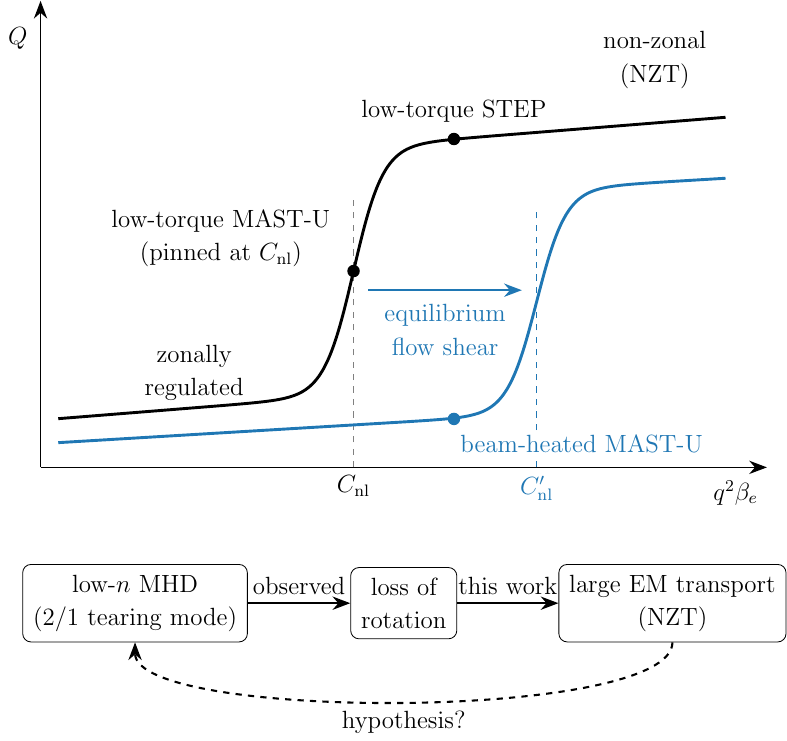}
  \caption{Top: schematic heat flux $Q$, on a logarithmic scale, against $q^2\beta_e$ without (black) and with (\textcolor{tabblue}{blue}) equilibrium flow shear. Dashed verticals, the transition constant without flow shear, $C_{\mathrm{nl}}$, and with it, $C_{\mathrm{nl}}'$. Filled circles, three operating points: a beam-heated MAST-U plasma between the two constants, in the zonally regulated state with flow shear (\textcolor{tabblue}{blue}, as at S50); the same point without flow shear, in the non-zonal state (black, low-torque STEP); and a plasma without momentum injection, whose profiles are pinned at $C_{\mathrm{nl}}$ by the stiff transport above it (black, low-torque MAST-U). NZT, the non-zonal transition. Bottom: the three possible causal links between large-scale (low-$n$) MHD, the loss of rotation and the non-zonal transition, each drawn as an arrow from cause to effect. Solid, the two established links (observed in \#48657, and this work); dashed, the hypothesised link that would close the loop.}
  \label{fig:nzt-cartoon}
\end{figure}

Figure~\ref{fig:nzt-cartoon} (bottom) sets out the three possible causal links between them, drawn as arrows from cause to effect. The first, from large-scale MHD to the loss of rotation, is observed: the $2/1$ mode brakes the plasma. The second, from the loss of rotation to the transition, is the result of this work, within the local model: once the shearing rate falls to roughly half its measured value, the smallest we have simulated, the turbulence crosses into the high-flux state and the field becomes stochastic, provided $q^2\beta_e$ remains above $C_{\mathrm{nl}}$. The ordering of events in \#48657 is consistent with this chain and not with its reverse: the rotation falls with the onset of the $2/1$ mode, and $\beta_N$ only afterwards (Figure~\ref{fig:mast-profiles}). In this sense the transition occurs \emph{if} the rotation is lost, although the large electromagnetic flux is inferred from the simulations, not observed. The third link, from the transition back to large-scale MHD, is a hypothesis that is not demonstrated here; could the box-scale, low-$n$ modes of the high-flux gyrokinetic simulations be how a large-scale MHD event appears in local gyrokinetics? To fully close this chain into a loop, the turbulence would have to brake the rotation that regulates it, as the transition would occur \emph{if and only if} the rotation is lost; demonstrating whether this full loop can be captured in a model will require a deeper study of momentum transport in electromagnetic turbulence. 

This is a very crucial question for STEP; a low-torque STEP, with no rotation to lose, may be susceptible to the transition from the outset (unless other actuators can be found to give access to higher confinement).

\subsection{Predictions for the next MAST-U campaign}
\label{sec:conclusions-predictions}

In beam-heated plasmas these three phenomena cannot be separated, since the heating and the torque arrive together; they can only be separated by heating without injecting momentum. A new electron Bernstein wave (EBW) heating system, capable of injecting several MW into MAST-U~\cite{Webster2023, Jacquet2025}, is planned to operate in the next campaign (MU06) alongside the existing and additional neutral beams, and a central aim of the programme will be to explore the roles of rotation and rotation shear on transport against $\beta_e$ in experiments that vary the beam and EBW power, and hence the net torque~\cite{harrison2026}. 

Our results make several predictions for MAST-U plasmas heated only using EBW. Firstly, gyrokinetics suggests that MAST-U EBW plasmas may not reach the performance of \#48657 without including momentum injection of NBI: with weak flow shear, the ion-scale electromagnetic turbulence would pin the profiles at $q^2\beta_e \approx C_{\mathrm{nl}}^{\mathrm{MAST}}$, so that $\beta_N$ at a given heating power would saturate below the beam-heated value, with a stiff response to additional heating; a dedicated study is in preparation~\cite{Zhang2026prep}. Secondly, the transport at that limit would be electromagnetic and ion-scale, carried by a stochastic field, with a large electron heat flux where the electron scale is stable~\cite{patel2026etg} and an appreciable ion heat flux, in contrast to the near-neoclassical ion transport of beam-heated H-modes~\cite{kaye2021}. High priority questions to be addressed in MAST-U EBW-only plasmas include:  can EBW-only MAST-U plasmas access high performance and can these plasmas get close to the electromagnetic transport threshold in $q^2 \beta_e$?  If MAST-U EBW plasmas access this threshold, it will be very interesting to see if adding beam torque at fixed heating power improves performance (due to flow-shear moving the transition threshold to a higher value). 

\subsection{Limitations}
\label{sec:conclusions-limitations}

The local gyrokinetic model does not describe the coherent low-$n$ MHD activity that is present over much of the discharge, and cannot determine its impact on turbulent transport or how the anomalous \texttt{TRANSP} flux is partitioned between turbulence and coherent modes. Hybrid magnetohydrodynamic--gyrokinetic or global gyrokinetic models, as noted in~\cite{Abel2013}, are required to describe low-$n$ ideal MHD perturbations and the turbulence evolving self-consistently. In the meantime, the EBW-heated plasmas of the next MAST-U campaign will offer an experimental route to study related questions.

\section*{Acknowledgements}

The authors would like to thank M. Abazorius, A. Bokshi, F. Casson, D. Dickinson, M. Giacomin, P.G. Ivanov, and F. Watts. This work has been
funded by STEP, a major technology and infrastructure programme led by UK Industrial Fusion Solutions Ltd (UKIFS),
which aims to deliver the UK’s prototype fusion power plant
and a path to the commercial viability of fusion. This work has been funded by the EPSRC Fusion Grant 2022/27 [grant number EP/W006839/1]. To obtain further information on the data and models underlying this paper please contact \url{PublicationsManager@ukaea.uk}. For the purpose of open access, the author(s) has applied a Creative Commons Attribution (CC BY) licence (where permitted by UKRI, `Open Government Licence' or `Creative Commons Attribution No-derivatives (CC BY-ND) licence' may be stated instead) to any Author Accepted Manuscript version arising. Part of this work was performed using resources provided by the Cambridge Service for Data Driven Discovery (CSD3) operated by the University of Cambridge Research Computing Service (\url{www.csd3.cam.ac.uk}), provided by Dell EMC and Intel using Tier-2 funding from the Engineering and Physical Sciences Research Council (capital grant EP/T022159/1), and DiRAC funding from the Science and Technology Facilities Council (\url{www.dirac.ac.uk}). Part of this work was performed using resources provided by the PITAGORA supercomputer of the National Supercomputing Consortium CINECA through EUROfusion.

\section*{References}
\bibliographystyle{unsrt}
\bibliography{bibliography}

\appendix

\section{Parallel phase-space discretisation: the Arakawa scheme}
\label{app:arakawa}

In the main text, we set \texttt{arakawa\_zv} to false, replacing the conservative Arakawa bracket in the parallel direction with centred differences. Centred differences add numerical damping along the field line, which removes an unphysical grid-scale mode, at the price of no longer conserving the gyrokinetic free energy exactly. A reasonable objection is that a scheme which no longer conserves the free energy might change the very physics we wish to test, here the balance between drive, zonal flows and dissipation that sets the non-zonal transition. The difficulty is not specific to \texttt{GENE}: any gyrokinetic code must damp grid-scale structure along the field line, and must do so carefully when electromagnetic fields feed back on it.\footnote{The upwind dissipation of \texttt{CGYRO}, for example, was recently changed for the same reason: it now conserves both density and parallel current exactly, removing a spurious grid-scale source that drove a slow finite-$\beta$ instability through the parallel current~\cite{gacode498}.} In this appendix we justify our choice by comparing three schemes: the conservative bracket (\texttt{arakawa\_zv} true), centred differences (\texttt{arakawa\_zv} false), and a third operator, $\tilde{J}$, which we construct here. $\tilde{J}$ is the bracket with one correction: on the field part of the distribution, where the bracket leaves a residual that drives the grid-scale mode through Amp\`ere's law, it is replaced by a term that removes the residual whilst keeping the free energy conserved exactly. $\tilde{J}$ is therefore conservative like the bracket, but free of its numerical source. Our findings are as follows.
\begin{itemize}
    \item At the longest wavelengths and on the parallel grids affordable in nonlinear simulations, the fastest-growing unstable mode (dominant) of the conservative bracket is not physical: it is a sawtooth that alternates in sign from one grid point to the next. The physical mode is still present, but as its second-fastest-growing mode (subdominant).
    \item The grid-scale mode is numerical. It disappears as the parallel grid is refined, and in the limit of sufficiently large $n_z$ all three schemes converge to the same physical mode and growth rate.
    \item Exact conservation does not guarantee accuracy. The bracket conserves the free energy to round-off, yet converges only on parallel grids too fine for nonlinear simulations, whereas centred differences converge smoothly.
    \item $\tilde{J}$ has no grid-scale mode at modest resolution, and its spectra agree with those of centred differences, so a conservative scheme and centred differences give the same physics once the numerical source is removed.
    \item Centred differences are therefore acceptable for this work. At the wavelengths that carry the transport, all three schemes agree to within a few per cent. Repeating the nonlinear $\beta_e$ scan at S50 with $\tilde{J}$ changes the state of no simulation, so the low-flux state and the location of the non-zonal transition do not depend on the parallel discretisation.
\end{itemize}
In the following, we explain how the grid-scale mode arises, construct $\tilde{J}$, and compare the three schemes linearly and nonlinearly.

\subsection{The parallel operator}
\label{app:arakawa-operator}

We denote by $z$ the position along a field line, by $v_\parallel$ the velocity along it, and by $s$ the particle species. The magnetic moment $\mu$ labels the perpendicular motion and is held fixed throughout. We write the distribution as an equilibrium Maxwellian $F_{0s}$ plus a small perturbation $f_s$. Parallel streaming moves it along $z$, and the mirror force changes $v_\parallel$ where the field strength $B(z)$ varies.

The effects due to parallel streaming and the mirror force can be written as a Poisson bracket. For any distribution $g$, we use the sign convention
\begin{equation}
  \{H_s,g\} \equiv
  \frac{\partial H_s}{\partial z}\frac{\partial g}{\partial v_\parallel}
  -\frac{\partial H_s}{\partial v_\parallel}\frac{\partial g}{\partial z}.
  \label{eq:app_bracket}
\end{equation}
Here, $H_s$ is the particle energy in thermal units, which we split into equilibrium and perturbed parts,
\begin{equation}
  H_s=H_{0s}+\Phi,\qquad H_{0s}=v_\parallel^2+\mu B(z).
  \label{eq:app_hamiltonian}
\end{equation}
The field term $\Phi$ is gyroaveraged and, at fixed $\mu$, does not depend on $v_\parallel$. In the linearised equation, the field term can either be evolved separately or combined with $f_s$ into a single bracket,
\begin{equation}
  h_s\equiv f_s+F_{0s}\Phi,\qquad
  \frac{\partial f_s}{\partial t}=\{H_{0s},h_s\}.
  \label{eq:app_h_form}
\end{equation}
The two forms are equal in the continuum through the chain and product rules, which finite differences do not satisfy exactly.

The bracket (\texttt{arakawa\_zv} true) applies the three-stencil average of~\cite{Arakawa1966} to $h_s$, chosen to preserve discrete conservation. Here, $J$ is the resulting equilibrium parallel operator in \texttt{GENE}, including its Maxwellian weighting. Its conservation property is antisymmetry in the free-energy inner product,
\begin{equation}
  \langle a, J b\rangle = -\langle J a, b\rangle ,
  \qquad
  \langle a, b\rangle \equiv \sum_{j,l} \frac{a_{j,l}^{*}\,b_{j,l}}{F_{0,j,l}} ,
  \label{eq:app_antisymmetry}
\end{equation}
where $j$ and $l$ index the $z$ and $v_\parallel$ grid points and the star denotes complex conjugation. We suppress the species index and the phase-space integration weights. If $\partial_t g=Jg$, then
\begin{equation}
  \frac{\mathrm{d}}{\mathrm{d}t}\langle g,g\rangle
  =\langle Jg,g\rangle+\langle g,Jg\rangle=0.
  \label{eq:app_energy_identity}
\end{equation}
so the parallel bracket produces no net free energy.

Centred differences (\texttt{arakawa\_zv} false) instead apply the difference to $f_s$ and add the field term pointwise,
\begin{equation}
  D(F_{0}\Phi)_{j,l} \equiv -\,2F_{0,j,l}\,v_l\,\delta_z[\Phi]_j ,
  \label{eq:app_dchidz}
\end{equation}
with $\delta_z$ the fourth-order centred difference and $v_l$ the parallel velocity at grid point $l$. This does not conserve the free energy exactly. The third choice, $\tilde{J}$, modifies $J$ so that it agrees with $D$ on the field part in the interior, whilst retaining antisymmetry. All three include fourth-order parallel hyperdiffusion, a numerical damping that is strongest on variations over a few grid points.

\subsection{Why the Arakawa bracket allows a mode at the grid scale}
\label{app:arakawa-grid-mode}

This subsection explains what the grid-scale mode of Figure~\ref{fig:arakawa} is, and why it grows with the Arakawa bracket but not with centred differences. It does so by identifying the properties of the mode and how each interacts with the numerical scheme. A numerical mode at the grid scale can only grow if three ingredients are present: it must have structure that the parallel derivative cannot see, a numerical source that feeds that structure, and a feedback through the fields that amplifies the source faster than hyperdiffusion damps it. We take each in turn, and state what it tells us.

Firstly, the parallel derivative cannot see the shortest structure on the grid. The shortest wavelength on the grid is the sawtooth $g_j=(-1)^j$, which alternates in sign from one point to the next. A centred stencil $Lg_j=\sum_u a_u g_{j+u}$, where $u$ is the integer offset from grid point $j$ and the coefficients are paired, $a_{-u}=-a_u$, returns zero on it,
\begin{equation}
  \sum_u a_u\,g_{j+u} = \sum_{u\ge1} a_u\left(g_{j+u} - g_{j-u}\right) = 0 ,
  \label{eq:app_sawtooth}
\end{equation}
since $g_{j+u} = g_{j-u}$ for the sawtooth. Increasing the order of the stencil does not remove this cancellation. Parallel streaming therefore neither moves nor disperses a sawtooth, and the only thing acting on it is the hyperdiffusion, which damps it. This tells us that the sawtooth is invisible to the physics of the parallel operator, and that it can grow only if something else feeds it faster than the hyperdiffusion removes it.

Secondly, the bracket feeds the parallel current at the grid scale, whereas centred differences do not. The bracket acts on $h_s = f_s + F_{0s}\Phi$ and so differentiates the field term as a product, whereas centred differences differentiate $\Phi$ alone and multiply by $F_0$ afterwards~(\ref{eq:app_dchidz}). In the continuum the two are equal by the chain and product rules, but on the grid they are not, and higher-order differences reduce the discrepancy for smooth functions without removing it. We denote the difference by
\begin{equation}
  \mathcal{R}(\Phi) \equiv J\!\left(F_{0}\Phi\right) - D\!\left(F_{0}\Phi\right).
  \label{eq:app_residual}
\end{equation}
This residual is odd in $v_\parallel$ on the symmetric velocity grid,
\begin{equation}
  \mathcal{R}(z,-v_\parallel)=-\mathcal{R}(z,v_\parallel),\qquad
  \sum_l \mathcal{R}_{j,l}=0,\qquad \sum_l v_l\mathcal{R}_{j,l}\ \hbox{need not be zero},
  \label{eq:app_residual_moments}
\end{equation}
so its contributions to the density at opposite velocities cancel, whilst its contributions to the parallel current add. Writing the bracket as
\begin{equation}
  Jh=Jf+D(F_0\Phi)+\mathcal{R}(\Phi)
  \label{eq:app_residual_split}
\end{equation}
makes the extra term explicit: the bracket is centred differences plus $\mathcal{R}$. This tells us where the numerical source lies. It is a spurious parallel current, present only with the bracket, which leaves the density untouched and so does not show up in the electrostatic response. Because $\mathcal{R}$ is a discretisation error, it shrinks as the parallel grid is refined, which is why the grid-scale mode disappears at large $n_z$.

Thirdly, Amp\`ere's law amplifies that current at long wavelength and high $\beta$. The perturbed current sets the parallel vector potential $A_\parallel$, which acts back on the distribution and so on the current itself. At small perpendicular wavenumber $\kperp$, the current and electromagnetic skin terms in the field equation are large and nearly cancel, leaving a small physical remainder~\cite{Pueschel2008}. A relative error $\eta$ in the current can then produce a relative frequency error of order
\begin{equation}
  \frac{\delta\omega}{\omega}\sim\frac{\eta}{\xi},
  \qquad \xi\equiv\frac{\kperp^2\rho_s^2}{\beta_e},
  \label{eq:app_current_amplification}
\end{equation}
where $\omega$ is the mode frequency and $\delta\omega$ its numerical error. This tells us when the source matters: the same current error is amplified by $1/\xi$, and so most strongly at the longest wavelengths and the highest $\beta_e$. That is where the bracket's dominant mode becomes the sawtooth in Figure~\ref{fig:arakawa}, whilst at shorter wavelengths all three schemes agree.

Together, the three ingredients explain the properties of the grid-scale mode. It is invisible to the parallel derivative, so nothing but hyperdiffusion opposes it; it is fed by the residual $\mathcal{R}$, which only the bracket contains; and it is amplified through $A_\parallel$ at long wavelength and high $\beta_e$. With centred differences, the field term is evaluated directly as $D(F_0\Phi)$, so the source is absent and hyperdiffusion holds the numerical mode close to marginal stability. A more physical remedy is therefore not to abandon conservation, but to remove $\mathcal{R}$ from the field part of the bracket whilst keeping its antisymmetry, which is what $\tilde{J}$ does.

\subsection{The operator $\tilde{J}$}
\label{app:arakawa-jtilde}

We now construct $\tilde{J}$, which must remove the residual on the field part whilst preserving the free-energy identity. Simply subtracting the residual does the first, but not necessarily the second. At each $z_j$, we take the part of a distribution with the velocity dependence of $F_0$,
\begin{equation}
  P h = F_{0}\,\psi_h ,
  \qquad
  \psi_{h,j} = \frac{\sum_l h_{j,l}}{\sum_l F_{0,j,l}}.
  \label{eq:app_proj}
\end{equation}
so that $Ph$ has the same velocity sum as $h$ but a Maxwellian shape. The operator that returns the residual is then
\begin{equation}
  T\equiv(J-D)MP,\qquad Th=\mathcal{R}(M\psi_h).
  \label{eq:app_correction}
\end{equation}
Here, $M$ multiplies by a real function of position, equal to one in the interior and falling smoothly to zero over the two end segments of the ballooning chain. Finally, we add the adjoint correction,
\begin{equation}
  \tilde{J} \equiv J - T + T^{\dagger}.
  \label{eq:app_skew}
\end{equation}
where the dagger denotes the adjoint in the inner product of~(\ref{eq:app_antisymmetry}). Since $J^\dagger=-J$, direct substitution gives
\begin{equation}
  \tilde{J}^{\dagger}
  =-J-T^{\dagger}+T=-\tilde{J}.
  \label{eq:app_skew_proof}
\end{equation}
so the cancellation in~(\ref{eq:app_energy_identity}) still holds. For any input $a$, $Ta$ is odd in $v_\parallel$, whereas $F_0\psi$ is even, so that $T^\dagger(F_0\psi)=0$. Where $M=1$ over the whole local stencil, we therefore obtain
\begin{equation}
  \tilde{J}(F_0\psi)
  =J(F_0\psi)-(J-D)(F_0\psi)=D(F_0\psi).
  \label{eq:app_corrected_field}
\end{equation}
which is the required discrete field term. We note that $\tilde{J}$ applies hyperdiffusion to $f_s$ rather than to $h_s$. This change alone leaves the grid-scale instability of the bracket essentially unchanged (not shown), so the improvement comes from correcting the parallel coupling, and not from additional hyperdiffusion.

\subsection{Numerical assessment}
\label{app:arakawa-assessment}

\begin{figure}
  \centering
  \includegraphics{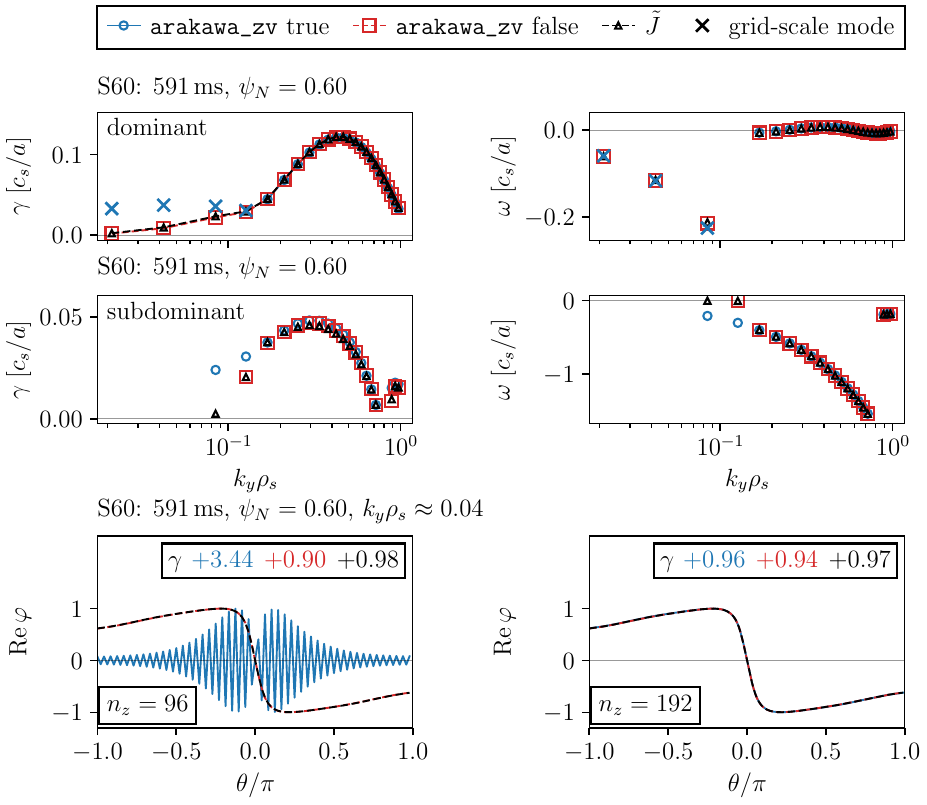}
  \caption{Growth rate (left-hand side) and real frequency (right-hand side) against $k_y\rho_s$ at S60, full-plasma $\beta'$, of each simulation's fastest-growing and second-fastest-growing unstable modes (dominant, top row; subdominant, middle row), by dynamic mode decomposition, for \texttt{arakawa\_zv} true (\textcolor{tabblue}{blue} circles), \texttt{arakawa\_zv} false (\textcolor{tabred}{red} squares, larger) and $\tilde{J}$ (black triangles, smaller), all open. A cross marks a grid-scale mode. Bottom row: $\mathrm{Re}\,\varphi$ of the fastest-growing unstable mode (dominant) at $k_y\rho_s \approx 0.04$ along one ballooning segment, normalised to $|\varphi|_{\max}$, with $n_z = 96$ (left-hand side) and $n_z = 192$ (right-hand side) parallel grid points, coloured as above; each panel gives each mode's growth rate in $10^{-2}\,c_s/a$.}
  \label{fig:arakawa}
\end{figure}

Figure~\ref{fig:arakawa} repeats the S60 $k_y$ scan of Figure~\ref{fig:spectra} with each scheme, extracting the two fastest-growing unstable modes by dynamic mode decomposition~\cite{Schmid2010,Dudkovskaia2025}. Above $k_y\rho_s = 0.17$, the three schemes agree to better than one per cent up to $k_y\rho_s \approx 0.7$, and to within about three per cent up to $k_y\rho_s \approx 1$. Below $k_y\rho_s = 0.17$, the fastest-growing unstable mode (dominant) of the bracket is the sawtooth, and the physical mode is its second-fastest-growing unstable mode (subdominant), whereas the other two schemes return the physical mode as dominant. The bottom row of Figure~\ref{fig:arakawa} shows the corresponding mode structures at $k_y\rho_s \approx 0.04$. On a coarse parallel grid, the dominant mode of the bracket is a sawtooth that alternates in sign from one grid point to the next, whilst centred differences and $\tilde{J}$ return the smooth physical mode. On a finer grid, the sawtooth is gone and all three schemes return the same physical mode.

\begin{figure}
  \centering
  \includegraphics{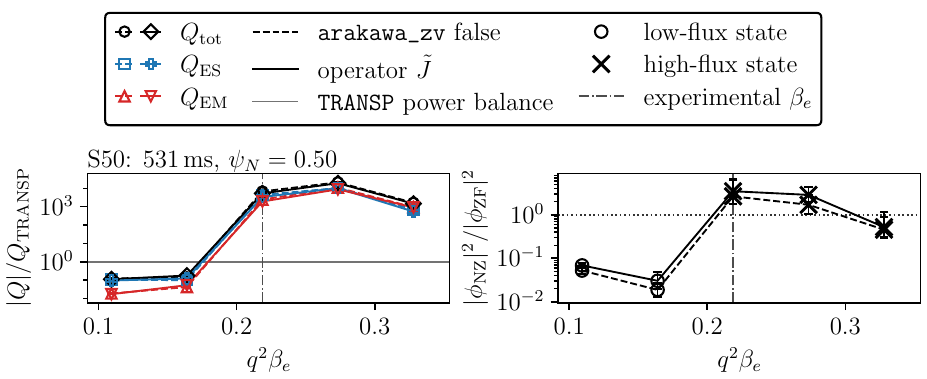}
  \caption{Nonlinear S50 simulations against $q^2\beta_e$ (dash-dot, experimental $\beta_e$) with \texttt{arakawa\_zv} false, as in Figure~\ref{fig:nonlinear-betascan} (dashed; circles, squares and triangles on the left-hand side), and with $\tilde{J}$ in single precision (solid; diamonds, plus signs and inverted triangles). All markers are open. $|Q|$ summed over species in units of $Q_{\mathrm{TRANSP}}$, coloured and with the grey line as in Figure~\ref{fig:nonlinear-betascan} (left-hand side). $|\varphi_{\mathrm{NZ}}|^2/|\varphi_{\mathrm{ZF}}|^2$ of~(\ref{eq:nz-zf}) with 16th to 84th percentile bars, circles low-flux and crosses high-flux in both (right-hand side). Each pair is averaged over the last fifth of the shorter simulation.}
  \label{fig:arakawa-nl}
\end{figure}

When the parallel grid is refined at $n_0 = 2$, the sawtooth is dominant for the bracket on the two coarsest grids, but it is never one of the two fastest-growing modes of $\tilde{J}$. The bracket conserves the free energy to round-off. However, at every $n_{v_\parallel}$, it converges only on parallel grids too fine for nonlinear simulations, so exact conservation does not guarantee accuracy. Centred differences converge smoothly, but on the coarsest grid they produce free energy comparable to the drive. $\tilde{J}$ conserves the free energy to round-off, and its growth rate varies by a few per cent over the grids, compared with about a tenth for centred differences. With $\tilde{J}$, no nonlinear S50 simulation of Figure~\ref{fig:nonlinear-betascan} changes state, and the low-flux fluxes agree to within a factor of two (Figure~\ref{fig:arakawa-nl}). The low-flux state and the location of the transition are therefore insensitive to the parallel discretisation.

We have tested $\tilde{J}$ only linearly at S60 and nonlinearly at S50. Centred differences return the physical mode at every wavenumber, with a growth rate slightly below the converged value on coarse parallel grids.

\FloatBarrier

\section{The collision operator}
\label{app:collisions}

\begin{figure}[tbp]
  \centering
  \includegraphics{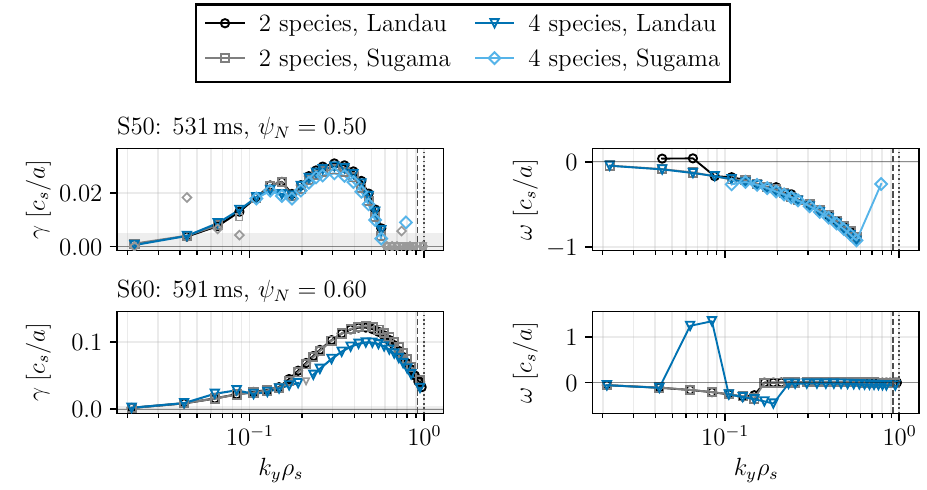}
  \caption{Growth rate (left-hand side) and real frequency, positive in the ion direction (right-hand side), of the fastest-growing unstable mode (dominant) against $k_y\rho_s$ at the full-plasma $\beta'$ without equilibrium flow, one row per surface, with \texttt{GENE}'s Landau and Sugama operators: two species, Landau (black circles) and Sugama (grey squares); four species, Landau (dark blue triangles) and, at S50 only, Sugama (light blue diamonds). Every marker is open; full-size markers in the curve colour are converged growing modes and small grey markers damped or unconverged modes. Grey band, dashed and dotted verticals as in Figure~\ref{fig:spectra}.}
  \label{fig:spectra-species}
\end{figure}

We use \texttt{GENE}'s linearised Landau operator with Xu--Rosenbluth conservation terms and finite-Larmor-radius corrections; Figure~\ref{fig:spectra-species} compares the two-species scans, and the four-species scan at S50, with the Sugama operator. At S50 the Sugama growth rate is lower by several per cent at most from $k_y\rho_s \approx 0.1$ up to the peak, most at the peak, with the frequencies unchanged. With four species it is lower there by under a tenth, again most at the peak. At S60 it is lower by between two and six per cent from $k_y\rho_s \approx 0.1$ to the peak and higher by a few per cent beyond it. Below $k_y\rho_s \approx 0.1$ the Sugama growth rates are much smaller on both surfaces, by an order of magnitude at the longest wavelengths. Between $k_y\rho_s \approx 0.1$ and the peak the choice of operator therefore changes the growth rate by less than a tenth. Beyond the S50 peak the absolute change stays comparable, but the relative change grows as the growth rate falls towards zero.

\FloatBarrier

\section{The four-species parameters}
\label{app:species-table}

Table~\ref{tab:species_inputs} gives the four-species \texttt{TRANSP} parameters of the two analysis surfaces, from which the two- and three-species sets of Sections~\ref{sec:linear-species} and~\ref{sec:species-three-four} are built.

\begin{table}[tbp]
  \caption{Species parameters of the four-species \texttt{GENE} inputs at the two analysis surfaces, from the same \texttt{TRANSP} profiles as Table~\ref{tab:gk_setup}. Here $Z_s$ is the charge number of species $s$, $m_s/m_i$ its mass over the thermal deuterium mass, $n_s/n_e$ and $T_s/T_e$ its density and temperature over the electron values, and $a/L_{n_s}$ and $a/L_{T_s}$ its normalised density and temperature gradients, with $a/L_X = -a\,\mathrm{d}\ln X/\mathrm{d}r$ and $a$ the minor radius. The last line of each block gives $Z_{\mathrm{eff}} = \sum_s Z_s^2 n_s/n_e$ over the ion species and the quasineutrality sums $\sum_s Z_s n_s/n_e$ and $\sum_s Z_s n_s (a/L_{n_s})/n_e$ over all species. The two- and three-species reductions of these inputs are~(\ref{eq:species_two}) and~(\ref{eq:species_three}). \texttt{GENE} takes one reference collisionality and computes the collision frequency of every pair of species from the charges, masses, densities and temperatures of the species evolved, so that $(a/c_s)\nu_{ei} = 0.577$ and $0.559$ at S50 and S60 with these inputs, $Z_{\mathrm{eff}}$ times the two-species values of Table~\ref{tab:gk_setup}.}
  \label{tab:species_inputs}
  \centering
  \renewcommand{\arraystretch}{1.3}
  \small
  \begin{tabular*}{\linewidth}{@{\extracolsep{\fill}}lcccccc@{}}
\toprule
species & $Z_s$ & $m_s/m_i$ & $n_s/n_e$ & $T_s/T_e$ & $a/L_{n_s}$ & $a/L_{T_s}$ \\
\midrule
\multicolumn{7}{l}{S50: $531\,\mathrm{ms}$, $\psi_N = 0.50$} \\
electrons ($e$) & $-1$ & $2.72 \times 10^{-4}$ & 1.0000 & 1.00 & 0.73 & 1.37 \\
thermal deuterium ($i$) & 1 & 1.000 & 0.8884 & 1.34 & 0.68 & 1.66 \\
carbon ($C$) & 6 & 6.003 & 0.0167 & 1.33 & 0.73 & 1.67 \\
fast beam deuterium ($f$) & 1 & 1.000 & 0.0116 & 33.37 & 4.51 & $-0.63$ \\
\multicolumn{7}{l}{$Z_{\mathrm{eff}} = 1.50$, \quad $\sum_s Z_s n_s / n_e = 0.0000$, \quad $\sum_s Z_s n_s\,(a/L_{n_s}) / n_e = 0.0000$} \\
\midrule
\multicolumn{7}{l}{S60: $591\,\mathrm{ms}$, $\psi_N = 0.60$} \\
electrons ($e$) & $-1$ & $2.72 \times 10^{-4}$ & 1.0000 & 1.00 & 1.20 & 1.77 \\
thermal deuterium ($i$) & 1 & 1.000 & 0.8924 & 1.17 & 1.16 & 0.76 \\
carbon ($C$) & 6 & 6.003 & 0.0167 & 1.17 & 1.20 & 0.79 \\
fast beam deuterium ($f$) & 1 & 1.000 & 0.0076 & 35.59 & 6.22 & $-0.63$ \\
\multicolumn{7}{l}{$Z_{\mathrm{eff}} = 1.50$, \quad $\sum_s Z_s n_s / n_e = 0.0000$, \quad $\sum_s Z_s n_s\,(a/L_{n_s}) / n_e = 0.0000$} \\
\bottomrule
\end{tabular*}

\end{table}

\FloatBarrier

\section{Floquet analysis of a sheared linear mode}
\label{app:floquet}

In this appendix, we define the Floquet-averaged growth rate $\gamma_{\mathrm{F}}$ used in Section~\ref{sec:equilibrium-linear}, which is similar to the effective growth rate defined in \cite{Roach2009}. Without flow shear, an initial-value simulation converges to a single exponentially growing mode; with $E\times B$ shear it does not, since advection by $\bm{v}_E = \gamma_{E\times B}\,x\,\mathbf{y}$, with $x$ the radial coordinate and $\mathbf{y}$ the binormal unit vector, drifts each radial wavenumber~\cite{maeyama2025floquet,Newton2010},
\begin{equation}
  k_x(t) = k_x(0) - \gamma_{E\times B}\,k_y\,t .
  \label{eq:app_kx_drift}
\end{equation}
The envelope centre $\theta_0 = k_x/(\hat{s}k_y)$ therefore drifts as well, and the operator, $2\pi$-periodic in $\theta_0$, repeats after
\begin{equation}
  T_{\mathrm{F}} = \frac{2\pi\hat{s}}{\gamma_{E\times B}} ,
  \label{eq:app_TF}
\end{equation}
which is a few tens of $a/c_s$ at S50 and approximately $100\,a/c_s$ at S60. Only the net growth over a whole period is therefore meaningful. By Floquet's theorem~\cite{maeyama2025floquet},
\begin{equation}
  \dot{\mathbf{X}} = A(t)\,\mathbf{X}, \quad A(t + T_{\mathrm{F}}) = A(t), \qquad \mathbf{X}(t) = \mathrm{e}^{\lambda_{\mathrm{F}} t}\, \mathbf{U}(t), \quad \mathbf{U}(t + T_{\mathrm{F}}) = \mathbf{U}(t),
  \label{eq:app_floquet}
\end{equation}
where $\mathbf{X}(t)$ is the state vector of the linear system, $A(t)$ is its $T_{\mathrm{F}}$-periodic operator, $\lambda_{\mathrm{F}} = \gamma_{\mathrm{F}} + \mathrm{i}\omega_{\mathrm{F}}$ is the Floquet exponent and $\mathbf{U}$ is the periodic envelope. After a transient, the mode with the largest $\mathrm{Re}\,\lambda_{\mathrm{F}}$ dominates. Differencing $\ln|\mathbf{X}|$ of~(\ref{eq:app_floquet}) over whole periods gives
\begin{equation}
  \ln|\mathbf{X}(t)| = \gamma_{\mathrm{F}}\,t + \ln|\mathbf{U}(t)|, \qquad
  \gamma_{\mathrm{F}} = \frac{\ln|\mathbf{X}(t_0 + N_{\mathrm{p}}T_{\mathrm{F}})| - \ln|\mathbf{X}(t_0)|}{N_{\mathrm{p}}\,T_{\mathrm{F}}},
  \label{eq:app_fit}
\end{equation}
where $N_{\mathrm{p}}$ is an integer number of periods and $t_0$ is the start of the fit. Across $N_{\mathrm{p}} = 4$ to $7$ (up to six at S60), the fitted growth rate varies by at most $5\times10^{-3}\,c_s/a$, so we attach no meaning to the sign of a smaller growth rate (grey band in Figure~\ref{fig:spectra}).

\FloatBarrier

\end{document}